\documentclass[aps,pra,twocolumn,superscriptaddress]{revtex4-1}
\usepackage{bm}
\usepackage{graphicx}
\usepackage{amssymb,amsmath,amsbsy,amsgen,amsfonts}
\usepackage{dcolumn}
\usepackage{amsthm}
\usepackage{mathrsfs}
\usepackage{latexsym}
\usepackage{array}
\usepackage{amstext}
\usepackage{epsfig}
\usepackage{epstopdf} 

\usepackage{color}
\usepackage{units}
\usepackage{calrsfs}
\usepackage{verbatim}
\DeclareMathAlphabet\mathbfcal{OMS}{cmsy}{b}{n}

\newcommand{\be}{\begin{equation}}
\newcommand{\ee}{\end{equation}}
\newcommand{\ba}{\begin{array}}
\newcommand{\ea}{\end{array}}
\newcommand{\bqa}{\begin{eqnarray}}
\newcommand{\eqa}{\end{eqnarray}}

\begin{document}

\title{Fluctuation-induced forces on an atom near a photonic topological material}

\author{ M\'ario G. Silveirinha}
\email{mario.silveirinha@co.it.pt}
\address{Instituto Superior T\'{e}cnico, University of Lisbon
    and Instituto de Telecomunica\c{c}\~{o}es, Torre Norte, Av. Rovisco
    Pais 1, Lisbon 1049-001, Portugal}
\address{Laboratoire Charles Coulomb (L2C), UMR 5221 CNRS-Universit\'{e} de Montpellier, F-34095 Montpellier, France}

\author{S. Ali Hassani Gangaraj}
\email{ali.gangaraj@gmail.com}
\address{School of Electrical and Computer Engineering, Cornell University, Ithaca, NY 14853, USA}

\author{George W. Hanson}
\email{george@uwm.edu}
\address{Department of Electrical Engineering, University of Wisconsin-Milwaukee, 3200 N. Cramer St., Milwaukee, Wisconsin 53211, USA}

\author{Mauro Antezza}
\email{mauro.antezza@umontpellier.fr}
\address{Laboratoire Charles Coulomb (L2C), UMR 5221 CNRS-Universit\'{e} de Montpellier, F-34095 Montpellier, France}
\address{Institut Universitaire de France, 1 rue Descartes, F-75231 Paris Cedex 05, France}

\date{\today}

\begin{abstract}
We theoretically study the Casimir-Polder force on an atom in a
arbitrary initial state  in a rather general electromagnetic
environment wherein the materials may have a nonreciprocal
bianisotropic dispersive response. It is shown that under the Markov
approximation the force has resonant and nonresonant contributions.
We obtain explicit expressions for the optical force both in terms
of the system Green function and of the electromagnetic modes. We
apply the theory to the particular case wherein a two-level system
interacts with a topological gyrotropic material, showing that the
nonreciprocity enables exotic light-matter interactions and the
opportunity to sculpt and tune the Casimir-Polder forces on the
nanoscale. With a quasi-static approximation, we obtain a simple
analytical expression for the optical force and unveil the crucial
role of surface plasmons in fluctuation induced forces. Finally, we
derive the Green function for a gyrotropic material half-space in
terms of a Sommerfeld integral.
\end{abstract}

\maketitle

\section{Introduction}

The Casimir-Polder force acting on atoms located close to the
surface of a material body is of longstanding and current interest
\cite{CasimirPolder,DLP, Marvin, Milonni, Agarwal, Ford,
OptF1,OptF2,arXivBuhmann, Sipe1, Sipe2, Henkel, Boustimi, Fichet,
B2015, neq_Mauro, Exp_Cornell}, and is of considerable practical
importance in a variety of physical, biological and chemical
processes. For planar surfaces, the normal component of the force
has been extensively investigated both theoretically and
experimentally \cite{Mauro_revsm}. There is a vast literature on
theoretical methods to calculate the force when the material
structures are conventional isotropic dispersive dielectrics
\cite{Milonni,Agarwal, Ford, OptF1,OptF2, Sipe1, Sipe2, Henkel,
Boustimi, Fichet}. Furthermore, the Casimir-Lifshitz interactions
between two macroscopic bodies with exotic electromagnetic responses
have also been discussed in a variety of scenarios \cite{Leonhardt,
Zhao, MarioCF3, Rosa, Yannopapas, MarioCF1, MarioCF2, Stas1, Stas2,
Grushin, arXivBuhmann}, but the majority of the works consider
planar geometries and that the system is in the ground state.
Indeed, it seems that the Casimir-Polder interaction between a
neutral atom and a generic environment with a complex (e.g.,
gyrotropic or bianisotropic) electromagnetic response has not been
fully addressed so far in the literature.

In this article, motivated by the recent interest in nonreciprocal
photonic platforms with topological properties
\cite{Haldane1,Haldane2, MIT1,MIT2,MIT3, Silv1sm,Silv2,AliMulti,
Silv3, Hanson_1sm}, we develop a theoretical formalism to
characterize the Casimir-Polder force acting on an atom prepared in
an arbitrary initial state in the vicinity of a arbitrary possibly
bianisotropic, inhomogeneous and nonreciprocal dispersive system. In
the general case, the optical force is written in terms of the
system Green function. Interestingly, we show that in the limit of
vanishing material loss the quantum force may be written as well in
terms of the electromagnetic modes of the system.

We apply the developed formalism to a two-level atom placed in the
vicinity of a topological gyrotropic material, e.g., a magnetically
biased plasma \cite{Silv1sm,Silv2,AliMulti}. Based on a simple
quasi-static approximation, we obtain explicit formulas for the
fluctuation induced force and highlight how by tuning the strength
of the nonreciprocal response it is possible to tailor the amplitude
of the lateral and normal components of the optical force.
Furthermore, our analysis reveals that the fluctuation induced force
is largely determined by the surface plasmon polaritons (SPPs). The
``exact'' quantum force is numerically computed using the Green
function for an gyrotropic half-space, which is also derived here.
It is shown that the quasi-static approximation agrees rather well
with the result obtained with the exact Green function. Moreover, in
the companion article \cite{subPRL}, the developed theory is used to
show that excited atoms may experience nonzero spontaneous lateral
forces when near a photonic topological insulator. Unlike previous
studies \cite{Z2013,C2014,B2015,Z2015,I2017}, in a topological
system the sign of the lateral optical force may be polarization and
orientation-independent and is tunable \cite{subPRL}.

The article is organized as follows. In Sect. \ref{AI}, we derive
the vacuum fluctuation induced Casimir-Polder force acting on an
atom in a generic electromagnetic environment. The effect of thermal
fluctuations is neglected and the Markov approximation is used to
solve the Heisenberg equations. For simplicity, the analysis is
focused on two-level systems, but we provide also the expression of
the force for the case of multi-level atoms. In Sect. \ref{SectTop},
we consider the scenario wherein the electromagnetic environment is
a topological material half-space. Assuming that the material has a
gyrotropic response (magnetized plasma), we characterize the edge
(SPP) modes supported by the system and obtain closed-form
expressions for the Casimir-Polder force under a quasi-static
approximation. In Sect. \ref{SectNumeric} we present a numerical
study that illustrates how by controlling the strength of the
biasing magnetic field it is possible to tailor the amplitude and in
some cases also the sign of the Casimir-Polder force. Finally, a
short summary of the main findings is given in Sect.
\ref{SectConcl}.

\section{Optical force}

\label{AI}

In this section, we prove that in a rather general context the
expectation of the optical force acting on a two-level atom can be
decomposed into a resonant term (${{\mathcal{F}}_{R,i}}$) and a
non-resonant term (${{\mathcal{F}}_{C,i}}$) as
\begin{equation}
{{\mathcal{F}}_i}\left( t \right) = \left\langle {{{\hat
{\mathcal{F}}}_i}} \right\rangle  = {\rho _{ee}}\left( t
\right){{\mathcal{F}}_{R,i}} + \left( {1 - 2{\rho _{ee}}\left( t
\right)} \right){{\mathcal{F}}_{C,i}}, \label{Force_gen}
\end{equation}%
with ${\rho _{ee}}\left( t \right)$ the probability of the atom
being in the excited state. The resonant component of the force is
determined by the system Green function $\bf G$ (a $6 \times 6$
tensor, see Appendix \ref{ApGreen}) evaluated at the two-level atom
transition frequency ($\omega_0$)
\begin{equation}
{{\mathcal{F}}_{R,i}} = 2\,{\mathop{\rm Re}\nolimits} \left\{
{{{{\bf{\tilde \gamma }}}^*} \cdot {{\left. {\left( { - i\omega
{\partial _i}{\bf{G}}\left( {{\bf{r}},{{\bf{r}}_0};\omega } \right)}
\right)} \right|}_{\scriptstyle\omega  = {\omega _0} + i{0^+
}\hfill\atop \scriptstyle{\bf{r}} = {{\bf{r}}_0}\hfill}} \cdot
{\bf{\tilde \gamma }}} \right\}. \label{Force_res}
\end{equation}
Here, ${\partial _i} = {{\bf{\hat u}}_i} \cdot {\nabla _{\bf{r}}}$
represents the spatial derivative along the $i$--th space direction,
${\bf{\tilde \gamma }} = {\left[ {\begin{array}{*{20}{c}}
{\bf{\gamma }}&0
\end{array}} \right]^T}$ is a six-vector and $\gamma$ is the dipole transition matrix
element. The atom coordinates are determined by the vector
${{\bf{r}}_0}$. It is assumed that the atom is surrounded by a
vacuum (free-space) in its immediate vicinity.

The non-resonant component of the force gives the Casimir-Polder
force due to the zero-point fluctuations ${{\bf{\mathcal{F}}}_{C}}=
- {\nabla _{{{\bf{r}}_0}}}{{{\mathcal{E}}_{C}}}$, and depends on the
interaction Casimir energy
\begin{equation} \label{ECasimir}
\mathcal{E}_{C} = \frac{{ - 1}}{{4\pi }}\int\limits_{ - \infty
}^\infty  {d\xi {\rm{ }}} {\rm{tr}}\left( {{\bf{\tilde \alpha
}}\left( {i\xi } \right) \cdot {{\left( { - i\omega {{\bf{G}} }}
\right)}_{\omega  = i\xi }}} \right).
\end{equation}
In the above, ``$\rm{tr}$'' stands for the trace of a matrix and
${\bf{\tilde \alpha }}(\omega) = \left( {\frac{1}{{{\omega _0} -
\omega }}{\bf{\tilde \gamma }}{{{\bf{\tilde \gamma }}}^*} +
\frac{1}{{{\omega _0} + \omega }}{{{\bf{\tilde \gamma
}}}^*}{\bf{\tilde \gamma }}} \right)$, so that ${\bf{ \alpha
}}_{ij}={\bf{\tilde \alpha }}_{ij}/(\hbar \varepsilon_0)$ represents
the semiclassical ground-state electric polarizability of the
two-level system ($i,j=1,2,3$) \cite{footnoteLoss}. The Green
function is evaluated at imaginary frequencies ($\omega = i\xi$)
with identical observation and source points, ${\bf{r}} = {\bf{r'}}
= {{\bf{r}}_0}$. The result (\ref{Force_gen}) holds in the low
temperature limit: ${k_B}T \ll \hbar {\omega _0}$ and $d \ll
\lambda_T$, with $d$ the minimum distance between the atom and the
macroscopic bodies and $\lambda_T = hc/{k_B}T$ the thermal
wavelength. The Green function can be generally decomposed as ${\bf
G} = {\bf G}_0 + {\bf G}_s$, with ${\bf G}_0$ the free-space Green
function corresponding to the situation wherein the atom resides in
a vacuum. Due to symmetry reasons, ${\bf G}_0$ cannot contribute to
the force in the electric dipole approximation. Hence, in Eqs.
(\ref{Force_res}) and (\ref{ECasimir}) the Green function can be
replaced by its ``scattering part'' ${\bf G}_s$, which is free of
singularities when ${\bf{r}} = {\bf{r'}} = {{\bf{r}}_0}$.

For a two-level system the excited state probability is ${\rho
_{ee}}\left( t \right) = {\rho _{ee}}\left( 0 \right){e^{ - {\Gamma
_{eg}}t}}$ with
\begin{equation} \label{decayrate}
{\Gamma _{eg}} = \frac{2}{\hbar }{\mathop{\rm Im}\nolimits} \left\{
{{{{\bf{\tilde \gamma }}}^*} \cdot {{\left. {\left( { - i\omega
{\bf{G}}} \right)} \right|}_{\omega  = {\omega _0}}} \cdot
{\bf{\tilde \gamma }}} \right\}
\end{equation} the standard
spontaneous emission decay rate \cite{Mario_movingsm}.

Note that in the electric dipole approximation the force only
depends on the ``electric part'' of the Green function,
$\boldsymbol{\mathrm{G}}_{\rm{EE}}$, defined as in Eq.
(\ref{Greendecomp}) of Appendix \ref{ApGreen} (a $3 \times 3$
tensor). For standard dielectric media (with a trivial magnetic
response and vanishing magneto-electric tensors)
$\boldsymbol{\mathrm{G}}_{\rm{EE}}$ is related to the more
conventional Green function definition $\mathbfcal{G}$ of Refs.
\cite{OptF1, OptF2} as ${{\bf{G}}_{{\rm{EE}}}} = i\omega {\mu _0}
\mathbfcal{G}$.

\subsection{Modal Expansion}

To begin with, we obtain a formula for the optical force in terms of
the natural modes of oscillation of the electromagnetic field.
Hence, in this section we consider the limit of vanishing material
loss. For convenience, we adopt six-vector notations so that the
quantized electromagnetic fields are denoted by the six-vector
operator $\hat{\mathbf{F}} = (\hat{\mathbf{E}} ~~
\hat{\mathbf{H}})^{\mathrm{T}} $. The hat indicates that a given
symbol represents an operator.

From the correspondence principle, the optical force operator is
(electric dipole approximation) \cite{CATsm}
\begin{equation}
\hat{\mathcal{F}}_{j}=\hat{\mathbf{p}}_{g}\cdot \frac{\partial }{\partial j}%
\hat{\mathbf{F}},~~~j=x,y,z
\end{equation}%
where $\hat{\mathbf{p}}_{g}=(\hat{\mathbf{p}}~~\hat{\mathbf{0}})^{\mathrm{T}%
}$ is a generalized dipole moment operator and $\hat{\mathbf{p}}$ is
the standard electric dipole operator for the two-level atom. The
quantized electromagnetic field in a generic inhomogeneous and
dispersive material platform can be written in terms of positive and negative frequency components $\hat{\mathbf{F}}=%
\hat{\mathbf{F}}_{-}+\hat{\mathbf{F}}_{+}$ with ${\mathbf{{\hat{F}}}_{+}}=%
\mathbf{{\hat{F}}}_{-}^{\dag }$, and \cite{Mario_movingsm,
HeatTransportsm, Sipesm, SilvModalExpansions}
\begin{equation}
\hat{\mathbf{F}}_{-}(\mathbf{r},t)=\sum_{\omega _{n\mathbf{k}}>0}\sqrt{\frac{%
\hbar \omega _{n\mathbf{k}}}{2}}{\mathbf{F}}_{n\mathbf{k}}(\mathbf{r})\hat{a}%
_{n\mathbf{k}}(t).  \label{F}
\end{equation}%
In the above, ${\mathbf{F}}_{n\mathbf{k}}(\mathbf{r})$ represents a
generic cavity mode with oscillation frequency ${n\mathbf{k}}$, and
$\hat{a}_{n\mathbf{k}}(t)$ is the corresponding bosonic
operator satisfying $[\hat{a}_{n\boldsymbol{\mathrm{k}}},\hat{a}_{n%
\boldsymbol{\mathrm{k}}}^{\dagger }]=1$. The electromagnetic modes ${%
\mathbf{F}}_{n\mathbf{k}}$ are normalized as \cite{Mario_movingsm,
HeatTransportsm, Sipesm, SilvModalExpansions},
\begin{equation}
\frac{1}{2}\int\limits_{{}}{{d^{3}}\mathbf{r}\,\mathbf{F}_{n\mathbf{k}%
}^{\ast }\cdot \frac{{\partial \left( {\omega \mathbf{M}}\right) }}{{%
\partial \omega }}\cdot \,}\mathbf{F}_{n\mathbf{k}}^{{}}=1,  \label{norm}
\end{equation}%
where $\mathbf{M}= \mathbf{M} \left( {{\bf{r}},\omega } \right)$ is
the $6\times 6$ material matrix that describes the electromagnetic
properties of the environment. It relates the classical $\bf D$ and
$\bf B$ fields with the classical $\bf E$ and $\bf H$ fields. For a
generic bianisotropic (eventually nonreciprocal) material it is of
the form
\begin{equation}{\bf{M}} \left( {{\bf{r}},\omega } \right) = \left( {\begin{array}{*{20}{c}}
{\boldsymbol{\varepsilon }}&{\frac{1}{c}{\boldsymbol{\xi }}}\\
{\frac{1}{c} \boldsymbol{\zeta} }&{\boldsymbol{\mu }}
\end{array}} \right).
\end{equation}
The 3 $\times$ 3 tensors ${\boldsymbol{\varepsilon }}$ and
${\boldsymbol{\mu }}$ represent the permittivity and permeability,
and the tensors ${\boldsymbol{\xi }}$ and ${\boldsymbol{\zeta }}$
determine the magneto-electric response.

Using normal ordering of the field operators, the expectation of the
force can be written as
\begin{equation}
\mathcal{F}_{j}=\left\langle \hat{\mathcal{F}_{j}}\right\rangle =2 \, \mathrm{Re}%
\left\langle \hat{\boldsymbol{\mathrm{p}}}_{g}\cdot \partial _{j}\hat{%
\boldsymbol{\mathrm{F}}}_{-}\right\rangle .\   \label{Force}
\end{equation}%
In the above, the field is evaluated at $\mathbf{r}=\mathbf{r}_{0}$,
the position of the atom, and the Heisenberg picture is implicit.

The total Hamiltonian of the system is
\begin{align}
& \hat{\mathrm{H}}=\hbar \omega _{0}\hat{\sigma}_{+}\hat{\sigma}%
_{-}+\sum_{\omega _{n\mathbf{k}}>0}\frac{\hbar \omega _{n\mathbf{k}}}{2}%
\left( \hat{a}_{n\mathbf{k}}\hat{a}_{n\mathbf{k}}^{\dagger }+\hat{a}_{n%
\mathbf{k}}^{\dagger }\hat{a}_{n\mathbf{k}}\right)  \nonumber \\
& ~~~~~-\hat{\mathbf{p}}\cdot \hat{\mathbf{E}}(\mathbf{{r}_{0}}),
\end{align}%
where the last term is the interaction Hamiltonian ${\hat{H}_{{%
\mathop{\rm int}}}}$. With  $\hat{\boldsymbol{\mathrm{p}}}={\boldsymbol{\gamma }}^{\ast }\hat{\sigma%
}_{+}+{\boldsymbol{\gamma }}\hat{\sigma}_{-}$, it can be written as
\begin{align}
{\hat{H}_{{\mathop{\rm int}}}}&=-\left( \boldsymbol{\gamma }^{\ast }\hat{%
\sigma}_{+}+\boldsymbol{\gamma }\hat{\sigma}_{-}\right) \cdot \hat{\mathbf{E}%
}(\mathbf{{r}_{0}}) \\ \nonumber
&=-\left( \tilde{\boldsymbol{\gamma }}^{\ast }\hat{\sigma}%
_{+}+\tilde{\boldsymbol{\gamma }}\hat{\sigma}_{-}\right) \cdot \hat{\mathbf{F%
}}(\mathbf{{r}_{0}}),
\end{align}%
where $\hat{\sigma}_{\pm }$ are the atom raising and lowering
operators.

Using the Heisenberg equation of motion, $\partial _{t}\hat{a}_{n\mathbf{k}%
}=i\hbar ^{-1}\left[ \hat{ {H}},\hat{a}_{n\mathbf{k}}\right] $, it
follows that
\begin{equation}
\frac{\partial \hat{a}_{n\mathbf{k}}}{\partial t}=-i\omega _{n\mathbf{k}}%
\hat{a}_{n\mathbf{k}}+\frac{i}{\hbar }\hat{\mathbf{p}}\cdot \sqrt{\frac{%
\hbar \omega _{n\mathbf{k}}}{2}}{\mathbf{E}_{{n\mathbf{k}}}^{\ast }}(\mathbf{%
r}_{0}).
\end{equation}%
By integrating the differential equation one obtains \cite%
{Mario_movingsm},
\begin{align}
& \hat{a}_{n\boldsymbol{\mathrm{k}}}(t)=\hat{a}_{n\boldsymbol{\mathrm{k}}%
}e^{-i\omega _{n\boldsymbol{\mathrm{k}}}t}  \nonumber \\
& +\int \frac{i}{\hbar }\hat{\boldsymbol{\mathrm{p}}}(t^{\prime
})\cdot
\sqrt{\frac{\hbar \omega _{n\boldsymbol{\mathrm{k}}}}{2}}\boldsymbol{\mathrm{%
E}}_{n\boldsymbol{\mathrm{k}}}^{\ast }(\boldsymbol{\mathrm{r}}%
_{0})u(t-t^{\prime })e^{-i\omega
_{n\boldsymbol{\mathrm{k}}}(t-t^{\prime })}dt^{\prime }.
\end{align}%
Using the Markov approximation and
\begin{align}
\int\limits_{t_{0}}^{t}u(t-t^{\prime })&e^{-i(\omega _{n\boldsymbol{\mathrm{%
k}}}-\omega _{0})(t-t^{\prime })}dt^{\prime } \\
& \approx \pi \delta (\omega _{n\boldsymbol{\mathrm{k}}}-\omega _{0})+\text{%
PV}\frac{1}{i\left( \omega _{n\mathbf{k}}-\omega _{0}\right) }
\nonumber
\end{align}%
for an interaction that starts at $t_{0}\rightarrow -\infty $, it is
found that
\begin{align}
\hat{a}_{n\mathbf{k}}(t)\approx & \, \hat{a}_{n\mathbf{k}}e^{-i\omega _{n%
\mathbf{k}}t}  \label{a_hat} \\
& +\sqrt{\frac{\omega _{n\mathbf{k}}}{2\hbar }}\tilde{\boldsymbol{\gamma }}%
\cdot {\mathbf{F}}_{n\mathbf{k}}^{\ast
}\hat{\sigma}_{-}(t)\frac{1}{\omega
_{n\mathbf{k}}-\omega _{0}-i0^{+}}  \nonumber \\
& +\sqrt{\frac{\omega _{n\mathbf{k}}}{2\hbar }}\tilde{\boldsymbol{\gamma }}%
^{\ast }\cdot {\mathbf{F}}_{n\mathbf{k}}^{\ast }\hat{\sigma}_{+}(t)\frac{1}{%
\omega _{n\mathbf{k}}+\omega _{0}-i0^{+}}  \nonumber
\end{align}%
in the sense of the Sokhotski--Plemelj relation $\left( x\pm
i0^{+}\right) ^{-1}=\ $PV$\left( 1/x\right) \mp i\pi \delta \left(
x\right)$ ($\rm{PV}$ stands for the principal value). Assuming that
the photon field is initially in the ground state and using
(\ref{F}) and (\ref{Force}) one obtains the desired modal expansion
for the optical force
\begin{equation}\label{FE1}
\mathcal{F}_{j}=\rho _{ee}(t)\Sigma _{1}+\left( 1-\rho
_{ee}(t)\right) \Sigma _{2}
\end{equation}%
where
\begin{align} \label{sigma}
\Sigma _{1}& =\mathrm{Re}\left( \sum_{\omega _{n\mathbf{k}}>0}\omega _{n%
\mathbf{k}}\tilde{\boldsymbol{\gamma }}^{\ast }\cdot \partial _{j}{\mathbf{F}%
}_{n\mathbf{k}}\otimes \mathbf{F}_{n\mathbf{k}}^{\ast }\cdot \tilde{%
\boldsymbol{\gamma }}\frac{1}{\omega _{n\mathbf{k}}-\omega
-i0^{+}}\right) ,
\nonumber \\
\Sigma _{2}& =\mathrm{Re}\left( \sum_{\omega _{n\mathbf{k}}>0}\omega _{n%
\mathbf{k}}\tilde{\boldsymbol{\gamma }}\cdot \partial _{j}{\mathbf{F}}_{n%
\mathbf{k}}\otimes \mathbf{F}_{n\mathbf{k}}^{\ast }\cdot \tilde{\boldsymbol{%
\gamma }}^{\ast }\frac{1}{\omega _{n\mathbf{k}}+\omega
-i0^{+}}\right).
\end{align}%
We introduced $\rho _{ee}(t)=\left\langle
\hat{\sigma}_{+}\hat{\sigma}_{-}\right\rangle $. which gives the
probability of the atom to be found in its excited state in a
spontaneous emission process.

\subsection{Green function representation}

In what follows, it is shown that the optical force can also be
expressed in terms of the Green function $\bf{G}$ of the system. The
Green function ${\bf{G}}={\bf{G}}\left( {{\bf{r}},{\bf{r'}},\omega }
\right)$ is a $6 \times 6$ tensor defined by Eq. (\ref{Green}) of Appendix \ref{ApGreen}. With the help of Eq. (\ref{Gmp}) one
may rewrite the optical force (\ref{FE1}) as
\begin{widetext}
\begin{equation}\label{WE1}
\mathcal{F}_{j}=2\rho _{ee}(t)\mathrm{Re}\left\{
\tilde{\boldsymbol{\gamma
}}^{\ast }\cdot (-i\omega \partial _{j})\mathbf{G}^{+}(\mathbf{r_{0}},%
\mathbf{r_{0}},\omega _{0}+i0^{+})\cdot \tilde{\boldsymbol{\gamma
}}\right\}
 +2\left( 1-\rho _{ee}(t)\right) \mathrm{Re}\left\{ \tilde{\boldsymbol{%
\gamma }}^{\ast} \cdot (-i\omega \partial _{j})\mathbf{G}^{-}(\mathbf{r_{0}},\mathbf{%
r_{0}},\omega _{0}+i0^{+})\cdot \tilde{\boldsymbol{\gamma }}\right\}
\end{equation}%
\end{widetext}
where ${\mathbf{G}}^{\pm}$ are the positive/negative frequency parts
of the Green function, and the spatial derivatives act only on the
first argument ($\bf r$) of the Green function. All the poles of
${\mathbf{G}}^{\pm}$ are in the positive/negative real frequency
axis, respectively.

From Appendix \ref{ApGreen}, we have ${\bf{G}} = {{\bf{G}}^ + } +
{{\bf{G}}^ - } + \frac{1}{{i\omega }}{\bf{M}}_\infty ^{ - 1}\delta
\left( {{\bf{r}} - {{\bf{r}}_0}} \right)$. The $\delta$-function
term does not contribute to the force because it is associated with
the self-field, and hence it is possible to do the replacement
${\bf{G}}^+ \to \bf{G}-{\bf{G}}^ - $ in (\ref{WE1}). This leads to
Eq. (\ref{Force_gen}), with the Casimir-Polder force in the ground
state given by
\begin{equation}
\mathcal{F}_{C,j} = {\rm{2}}\,{\mathop{\rm Re}\nolimits} \left\{
{{{{\bf{\tilde \gamma }}}^*} \cdot \left( { - i\omega {\partial _j}}
\right){{\left. {{{\bf{G}}^ - }} \right|}_{\omega  = {\omega _0}}}
\cdot {\bf{\tilde \gamma }}} \right\}.
\end{equation}

Noting that ${{\bf{G}}^ - }$ is analytic for ${\mathop{\rm
Re}\nolimits} \left\{ \omega  \right\} > 0$, the Cauchy theorem
allows us to write the force as an integral over the imaginary
frequency axis,
\begin{equation}
\mathcal{F}_{C,j} = 2\,{\mathop{\rm Re}\nolimits} \left\{
{\frac{1}{{2\pi }}\int\limits_{ - \infty }^\infty  {d\xi }
{{\bf{\tilde \gamma }}^*} \cdot \frac{{{{\left( { - i\omega
{\partial _j}{{ {\bf{G}} }^ - }} \right)}_{\omega  = i\xi
}}}}{{{\omega _0} - i\xi }} \cdot {\bf{\tilde \gamma }}} \right\}.
\end{equation}
From the identity ${\left[ {\left( { - i\omega {\partial _j}}
\right){{\bf{G}}^ - }\left( {{\bf{r}},{{\bf{r}}_0}} \right)}
\right]^\dag }_{\scriptstyle{\bf{r}} = {{\bf{r}}_0}\hfill\atop
\scriptstyle\omega \hfill} = {\left[ {\left( { - i\omega {\partial
_j}} \right){{\bf{G}}^ - }\left( {{{\bf{r}}_0},{\bf{r}}} \right)}
\right]^{}}_{\scriptstyle{\bf{r}} = {{\bf{r}}_0}\hfill\atop
\scriptstyle{\omega ^*}\hfill}$, it follows that
${{\bf{\mathcal{F}}}_{C}}= - {\nabla
_{{{\bf{r}}_0}}}{{{\mathcal{E}}_{C}}}$, with the zero-point
interaction energy given by
\begin{equation}
\mathcal{E}_{C} = \frac{{ - 1}}{{2\pi }}\int\limits_{ - \infty
}^\infty  {d\xi } \frac{1}{{{\omega _0} - i\xi }}{{\bf{\tilde \gamma
}}^*} \cdot {\left( { - i\omega {{\bf{G}}^ - }} \right)_{\omega  =
i\xi }} \cdot {\bf{\tilde \gamma }}.
\end{equation}
Using again the analytic properties of ${{\bf{G}}^ - }$, we see that
$0 = \frac{1}{{2\pi }}\int\limits_{ - \infty }^\infty  {d\xi
\frac{1}{{{\omega _0} + i\xi }}} {\left( { - i\omega {{\bf{G}}^ - }}
\right)_{\omega  = i\xi }}$. Thus, introducing the tensor
${\bf{\tilde \alpha }} = \left( {\frac{1}{{{\omega _0} - \omega
}}{\bf{\tilde \gamma }}{{{\bf{\tilde \gamma }}}^*} +
\frac{1}{{{\omega _0} + \omega }}{{{\bf{\tilde \gamma
}}}^*}{\bf{\tilde \gamma }}} \right)$, which corresponds to a
normalized polarizability of the two-level atom, it is possible to
write
\begin{equation} \label{EC1}
\mathcal{E}_{C} = \frac{{ - 1}}{{2\pi }}\int\limits_{ - \infty
}^\infty  {d\xi {\rm{ }}} {\rm{tr}}\left( {{\bf{\tilde \alpha
}}\left( {i\xi } \right) \cdot {{\left( { - i\omega {{\bf{G}}^ - }}
\right)}_{\omega  = i\xi }}} \right),
\end{equation}
Noting finally that ${\left( { - i\omega {{\overline {\bf{G}} }^ -
}\left( {{{\bf{r}}_0},{{\bf{r}}_0},\omega } \right)} \right)_{\omega
= i\xi }} = {\left[ {{{\left( { - i\omega {{\overline {\bf{G}} }^ +
}\left( {{{\bf{r}}_0},{{\bf{r}}_0},\omega } \right)}
\right)}_{\omega  = i\xi }}} \right]^*}$ and that ${\bf{\tilde
\alpha }}\left( {i\xi } \right) = {{\bf{\tilde \alpha }}^*}\left(
{i\xi } \right)$, and taking into account that the integral is
necessarily real-valued, we see that the interaction Casimir energy
may be calculated using (\ref{EC1}) with ${{\bf{G}}^ + }$ in the
place of ${{\bf{G}}^ - }$. This result also implies that we can
replace ${{\bf{G}}^ - }$ by one-half of the full Green function
${{\bf{G}}}/2$ in (\ref{EC1}), and this final observation yields the
desired Eq. (\ref{ECasimir}), which may be written as
\begin{align}
{{\mathcal{E}}_{C}} =- \frac{1}{{4\pi }}\int\limits_{ - \infty
}^\infty  {d\xi {\rm{ }}} \left( {\frac{1}{{{\omega _0} - i\xi
}}{{{\bf{\tilde \gamma }}}^*} \cdot {{\left( { - i\omega
 {\bf{G}} } \right)}_{\omega  = i\xi }} \cdot {\bf{\tilde
\gamma }} + } \right. \nonumber \\
\left. {\frac{1}{{{\omega _0} + i\xi }}{\bf{\tilde \gamma }} \cdot
{{\left( { - i\omega  {\bf{G}} } \right)}_{\omega  = i\xi }} \cdot
{{{\bf{\tilde \gamma }}}^*}} \right).
\end{align}%

This completes the proof of equations
(\ref{Force_gen})-(\ref{ECasimir}). Even though the derivation
assumes negligible material loss, the final result is given in terms
of the Green function, and thus it can be readily extended to lossy
material systems simply by using the Green function for lossy
systems in the same expression. As previously mentioned, only the
scattering part of the Green function needs to be considered in the
force calculation, because by symmetry the self-field (i.e., the
part of the Green function associated with the radiation of an
electric dipole in a vacuum) does not contribute to the force.

Equation (\ref{Force_gen}) generalizes (in the low-temperature
limit) the theory of Refs. \cite{OptF1, OptF2} (which applies only
to isotropic dielectrics) to arbitrary bianisotropic (reciprocal or
nonreciprocal) material platforms. Note that different from Refs.
\cite{OptF1, OptF2} our theory neglects atomic level shifts and
broadenings. It is worth pointing out that for reciprocal systems
the integral in Eq. (\ref{ECasimir}) can be reduced to the positive
imaginary axis, but for general nonreciprocal systems the
integration must be over the entire imaginary axis.

\subsection{Lateral force for stratified systems}

\label{SubSectLateral}

So far the analysis is completely general (under the electric dipole
approximation), and applies to a generic system with no particular
symmetries. Next, we focus on structures invariant to translations
along the coordinates $\alpha =x,y$, and discuss some properties of
the optical lateral force in such systems.

Clearly, for a structure invariant to translations along $\alpha
=x,y$ the force component ${{\mathcal{F}}_{C,\alpha}}$ vanishes.
Furthermore, in the limit of no material loss we find from
(\ref{sigma}) with the Sokhotski--Plemelj relation, and using the
fact that the modes are Bloch waves, that
\begin{align}
& \mathcal{F}_{\alpha }=\rho _{ee}(t)\times   \nonumber \\
& \mathrm{Re}\left( i\pi \sum_{\omega _{n\mathbf{k}}>0}\omega _{n\mathbf{k}}%
\tilde{\boldsymbol{\gamma }}^{\ast }\cdot \partial _{\alpha }{\mathbf{F}}_{n%
\mathbf{k}}\otimes \mathbf{F}_{n\mathbf{k}}^{\ast }\cdot \tilde{\boldsymbol{%
\gamma }}\delta (\omega _{n\mathbf{k}}-\omega _{0})\right).
\end{align}%

In general, for lossy materials, the modal expansion does not apply
and one needs to use Eq. (\ref{Force_res}). As mentioned, the force
only depends on the ``electric part'' of the Green function
$\boldsymbol{\mathrm{G}}_{\rm{EE}}$. The vector
$\mathrm{\mathbf{E}}%
=-i\omega \boldsymbol{\mathrm{G}}_{\rm{EE}} \cdot
{\boldsymbol{\gamma }}$ corresponds to the frequency domain electric
field radiated by a classical dipole with electric dipole moment
$\boldsymbol{\gamma }$. The exact lateral force can be written in
terms of this electric field as follows (only the scattering part of
the field needs to be considered),
\begin{equation} \label{force_Efield}
\mathcal{F}_{\alpha }=2\rho _{ee}(t)
\mathrm{Re}\left\{ {\boldsymbol{\gamma }}^{\ast} \cdot \partial _{\alpha }%
 \bf{E}(\mathbf{r_{0}})\right\}.
\end{equation}%
The application of these formulas is illustrated in the companion
article \cite{subPRL}.

\subsection{Multi-level atom}

The formalism developed in the previous sections can be readily
generalized to a multi-level atom described by the Hamiltonian
${{\hat H}_{{\rm{at}}}} = \sum\limits_n {{E_n}\left| n \right\rangle
\left\langle n \right|}$, with $E_n$ the energy level of the $n$-th
state. It is supposed that the dipole moment matrix
${{\boldsymbol{\gamma }}_{mn}}=\left\langle {m|{\bf{\hat p}}|n}
\right\rangle$ has no diagonal elements. Thus, it is possible to
write the dipole moment operator as (for simplicity it is assumed
there are no degenerate levels)
\begin{equation}
{\bf{\hat p}} = \sum\limits_{{E_m} < {E_n}} {\left(
{{\boldsymbol{\gamma }}_{mn}^*\left| n \right\rangle \left\langle m
\right| + {{\boldsymbol{\gamma }}_{mn}}\left| m \right\rangle
\left\langle n \right|} \right)},
\end{equation}
which may be understood as a combination of multiple two-level
systems. Because Maxwell's equations are linear, equation
(\ref{a_hat}) can be readily generalized to a multi-level system by
including the contribution of each ``two-level'' term. Then,
substituting this result into (\ref{Force}) one sees that since the
Heisenberg equations preserve the orthogonality relations,
$\left\langle {m\left( t \right)|n\left( t \right)} \right\rangle  =
{\delta _{m,n}}$, each ``two-level'' component of the atom
Hamiltonian contributes independently to the force. Note that we
assume that in the initial state the electromagnetic field has no
quanta. This result proves that the optical force is a superposition
of the individual ``two-level'' contributions:
\begin{equation}
{{\mathcal{F}}_i}\left( t \right) = \sum\limits_{{E_m} < {E_n}} {
{\rho _{nn}}\left( t \right){{\mathcal{F}}^{mn}_{R,i}} + \left(
{\rho _{mm}\left( t \right) - {\rho _{nn}}\left( t \right)}
\right){{\mathcal{F}}^{mn}_{C,i}} }.
\end{equation}%
Here, ${\rho _{nn}}\left( t \right)$ is the probability of finding
the atom in the $n$-th state at time $t$, and
${{\mathcal{F}}^{mn}_{R,i}}$ and ${{\mathcal{F}}^{mn}_{C,i}}$ are
calculated using equations (\ref{Force_res})-(\ref{ECasimir}) with
${{\boldsymbol{\gamma }}_{mn}}$ in the place of
${{\boldsymbol{\gamma }}}$ and $\omega_{0,mn}=(E_n-E_m)/\hbar$ in
the place of $\omega_0$.

\section{Topological material half-space}

\label{SectTop}

\begin{figure}[tbh]
\begin{center}
\noindent \includegraphics[width=2.9in]{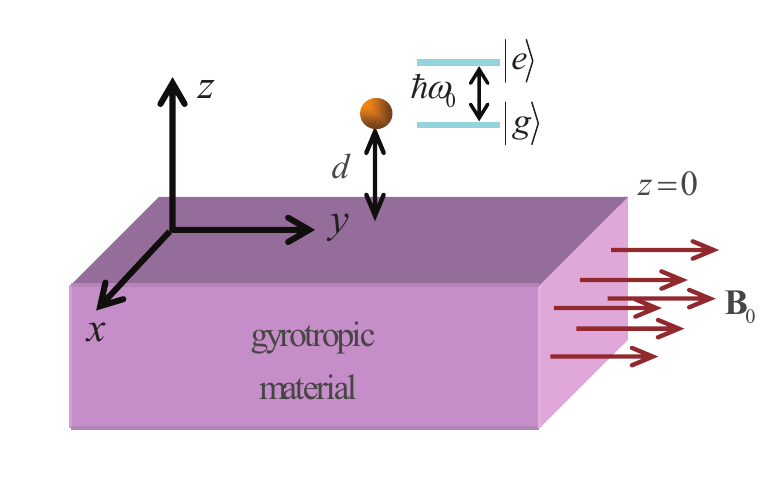}
\end{center}
\caption{A two-level system is at a distance $d$ above a gyrotropic
material. } \label{figatom}
\end{figure}

In the rest of the article, we focus on a $z$-stratified structure
formed by a topological material half-space ($z<0$) and a free-space
half-space ($z>0$) (Fig. \ref{figatom}). The atom is located a
distance $d$ above the topological material. It is assumed that the
material only has a
nontrivial electric response, so that $\mathbf{M}=\left( {%
\begin{array}{ccccccccccccccccccc}
{\boldsymbol{\varepsilon }\left({\mathbf{r},\omega }\right) } & 0 \\
0 & { {\mu _{0}} \boldsymbol{\rm{I}}}
\end{array}%
}\right) $. Furthermore, we suppose that the material response is
gyrotropic with dielectric function
\begin{equation}
{\boldsymbol{\varepsilon }}=\varepsilon _{0}(\varepsilon _{t}{\boldsymbol{%
\mathrm{I}}}_{t}+\varepsilon _{a}\mathbf{{\hat{y}}}\mathbf{{\hat{y}}}%
+i\varepsilon _{g}\mathbf{{\hat{y}}}\times \boldsymbol{\mathrm{I}}),
\label{permittivity}
\end{equation}%
where ${\boldsymbol{\mathrm{I}}}_{t}=\boldsymbol{\mathrm{I}}-\mathbf{{%
\hat{y}}{\hat{y}}}$ and $\varepsilon_g$ determines the strength of
the nonreciprocal response.

In Appendix \ref{ApGreenHalfSpace}, we derive an explicit formula
for the ``electric part'' of the Green function
$\boldsymbol{\mathrm{G}}_{\rm{EE}}$ in the region $z>0$ and $z'>0$.
The Green function has the decomposition ${{\bf{G}}_{{\rm{EE}}}} =
{{\bf{G}}_{{\rm{EE}}{\rm{,0}}}} + {{\bf{G}}_{{\rm{EE}}{\rm{,s}}}}$,
with ${{\bf{G}}_{{\rm{EE}}{\rm{,0}}}}$ the free-space Green function
(associated with the self-field) given by $\left( { - i\omega
{\varepsilon _0}} \right){{\bf{G}}_{{\rm{EE}}{\rm{,0}}}} = \left(
{\nabla \nabla  + k_0^2{\bf{\rm{I}}}} \right){\Phi _0}$ where ${\Phi
_0} = {{{e^{i{k_0}r}}} \mathord{\left/
 {\vphantom {{{e^{i{k_0}r}}} {4\pi r}}} \right.
 \kern-\nulldelimiterspace} {4\pi r}}$. The
scattering part of the Green function,
${{\bf{G}}_{{\rm{EE}}{\rm{,s}}}}$ is given by a Sommerfeld-type
integral
\begin{align}
\label{GEE} &(-i \omega
\varepsilon_0)\boldsymbol{\mathrm{G}}_{{\rm{EE}},s} \left(
{{\bf{r}},{\bf{r'}}} \right)= \nonumber \\
& \frac{1}{(2\pi )^{2}} \int \int d
{k}_{x}d {k}_{y} \, \frac{e^{-\gamma _{0}(z+z')}}{%
2\gamma _{0}}\, e^{i\mathbf{k}_{\Vert }\cdot
(\mathbf{r}-\mathbf{r}')} \, \mathbf{C}\left( \omega
,\mathbf{k}_{\Vert }\right)
\end{align}
where $\boldsymbol{\mathrm{k}}_{\parallel
}={k}_{x}\mathbf{\hat{x}}+{k}_{y}\mathbf{\hat{y}}$, ${\gamma _0} =
\sqrt {k_{\parallel}^2 - {k_0^2}}$, $k_0 = \omega/c$, and
$\mathbf{C}\left( \omega ,\mathbf{k}_{\Vert }\right)$ is the tensor
defined by Eq. (\ref{CintAp}), which is written in terms of the
reflection matrix for the gyrotropic material half-space.  To the best of our
knowledge, this is the first time that the Green function of a gyrotropic half-space
 is determined explicitly as a Sommerfeld integral.

By substituting (\ref{GEE}) into (\ref{Force_gen})-(\ref{ECasimir})
one obtains the exact solution for the optical force within the
Markov approximation for a transition between an excited state and
the ground. For example, the resonant component of the force can be
written as
\begin{equation} {{\mathcal{F}}_{R,i}} = 2\,{\mathop{\rm
Re}\nolimits} \left\{ {{{{\bf{ \gamma }}}^*} \cdot {{\left. {\left(
{ - i\omega {\partial _i}{\bf{G}}_{\rm EE}\left(
{{\bf{r}},{{\bf{r}}_0};\omega } \right)} \right)}
\right|}_{\scriptstyle\omega  = {\omega _0} + i{0^ + }\hfill\atop
\scriptstyle{\bf{r}} = {{\bf{r}}_0}\hfill}} \cdot {\bf{ \gamma }}}
\right\},
\end{equation}
with ${{\bf{r}}_0} = \left( {0,0,d} \right)$.

For a lossy magnetized plasma with bias magnetic field along the
$+y$-axis the permittivity elements are \cite{Bittencourt}
\begin{align}
\label{bp} &
{\varepsilon _t} = 1 - \frac{{\omega _p^2\left( {1 + i\Gamma /\omega } \right)}}{{{{\left( {\omega  + i\Gamma } \right)}^2} - \omega _c^2}} \nonumber \\
& {\varepsilon _a} = 1 - \frac{{\omega _p^2}}{{\omega \left( {\omega
+ i\Gamma } \right)}}, \,\,\,\, {\varepsilon _g} = \frac{1}{\omega
}\frac{{\omega _c^{}\omega _p^2}}{{\omega _c^2 - {{\left( {\omega  +
i\Gamma } \right)}^2}}}.
\end{align}
Here, $\omega _{p}$ is the plasma frequency, $\Gamma$ is the
collision rate associated with damping, $\omega _{c}=-qB_{0}/m$ is
the cyclotron frequency, $q=-e$ is the electron charge, $m$ is the
effective electron mass, and $B_{0}$ is the static bias. The
cyclotron frequency is either positive or negative depending if
$B_{0}$ is oriented along the $+y$ or $-y$ direction, respectively.
Narrow gap semiconductors such as InSb have a response analogous to
(\ref{bp}) \cite{Palik, GarciaVidal}.

It has been recently shown that electromagnetic continua with no
intrinsic periodicity but with broken time-reversal symmetry, e.g.,
the biased plasma described by (\ref{bp}), can be understood as
topological materials. In particular, such materials enable the
propagation of unidirectional, topologically protected and
scattering-immune edge states \cite{Silv1sm,Silv2,AliMulti,
HeatTransportsm}.

\begin{figure}[tb!]
\begin{center}
\noindent \includegraphics[width=3.5in]{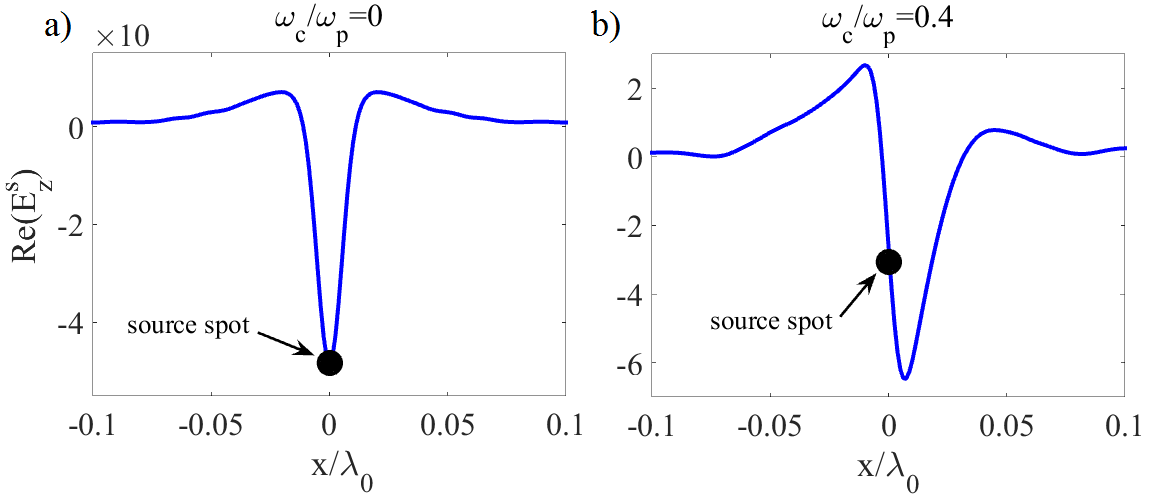}
\end{center}
\caption{Real part of the scattered electric field (in arbitrary
units) near the source for the unbiased plasma (a) and the biased
plasma (b). The oscillation frequency is  $\protect\omega /
\protect\omega_p = 0.7 $ and $\lambda_0 = 2 \pi c/\omega$. The
source (black dot) is located at $d = 0.05 c/\protect\omega_p $
above the magnetized plasma.} \label{figslope}
\end{figure}

From (\ref{Force_res}) and (\ref{force_Efield}) it is seen that the
lateral force is determined by the slope of the Green
function/electric field at the atom position \cite{subPRL}. To
illustrate that the slope is nonzero in the nonreciprocal case, we
consider that the dipole is polarized along the vertical ($z$)
direction (${\bf{E}} =  - i\omega {{\bf{G}}_{{\rm{EE}}}} \cdot
\gamma {\bf{\hat z}}$). As shown in Fig. \ref{figslope}a, for an
unbiased plasma (when $\omega_c=0$ and the dielectric function
reduces to the Drude dispersion scalar model), $\partial
_{x}E_{z}=0$ at the source point, and therefore $\mathcal{F}_{x}=0$.
However, in the
presence of a magnetic bias the field at the atom position has non-zero slope (Fig. \ref%
{figslope}b), and hence $\mathcal{F}_{x}\neq 0$. Note that Fig.
\ref{figslope} shows only the scattered part of the field at the
source point.

\section{Quasi-static solution}

To have some physical insight into the mechanisms that determine the
optical force, next we obtain an explicit expression for the force
under the assumption that $d \ll 2 \pi c/\omega_p$ and $d \ll 2 \pi
c/\omega_0$ (quasi-static limit) and that the material absorption is
negligible.

\subsection{Surface plasmon polaritons}

When the atom is in close vicinity of the topological material, the
light-matter interactions are expected to be mainly determined by
the surface plasmon polaritons (SPPs). In the following, we derive
the dispersion of the SPP resonances (i.e., SPPs with short
wavelengths and wave vector ${\bf k}_{\parallel } \to \infty$) based
on the quasi-static approximation. The propagation of SPPs has been
widely discussed in the literature mainly when the direction of
propagation is perpendicular to the bias field (e.g., \cite{Silv1sm,
Camley, EnghetaVoigt}), but it seems that for oblique directions the
problem was not systematically studied so far.

It is well known that SPPs with short wavelengths have an
electrostatic nature. Thus, we look for guided modes of the form
$\boldsymbol{\mathrm{F}}_{n\mathbf{k}}\approx \left[ \boldsymbol{\mathrm{E}}_{n\mathbf{k}}~~\boldsymbol{%
0}\right] ^{\mathrm{T}}\approx \left[ -\nabla \phi _{\mathbf{k}}~~%
\boldsymbol{0}\right] ^{\mathrm{T}}$. The magnetic field is assumed
negligible and the electric field is written in terms of
an electric potential ($\phi _{\mathbf{k}}$) that satisfies $\nabla \cdot (%
\boldsymbol{\varepsilon }\cdot \nabla \phi _{\mathbf{k}})=0$. The
solutions of this quasi-static equation are of the form
\begin{equation}
\phi _{\mathbf{k}}=\frac{{A}_{\mathbf{{k}_{\parallel }}}}{\sqrt{S}}%
e^{i\mathbf{{k}_{\parallel }}\cdot \boldsymbol{\mathrm{r}}}%
\begin{cases}
e^{-{k}_{\parallel }z}, & z>0 \\
e^{+\tilde{{k}}_{\parallel }z}, & z<0%
\end{cases}
\label{elecpot2}
\end{equation}%
where ${{\bf{k}}_{\parallel}} = {k_x}{\bf{\hat x}} + {k_y}{\bf{\hat
y}}$ is the wave vector of the SPPs,
${A}_{\mathbf{{k}_{\parallel}}}$ is a normalization parameter,
$\tilde{{k}}_{\parallel }=\sqrt{{k}_{x}^{2}+(\varepsilon
_{a}/\varepsilon _{t}){k}_{y}^{2}}$ and $S$ is the area of the slab.

Imposing that the normal component of the electric displacement is
continuous at the interface, i.e., that $\mathbf{\hat{z}}\cdot \mathbf{%
\varepsilon }\cdot \nabla {\phi _{\mathbf{k}}}$ is continuous at
$z=0$, we obtain the condition for the SPP resonance,
\begin{equation}\label{SPPresonance}
-{k}_{\parallel }={k}_{x}\varepsilon _{g}(\omega )+\tilde{%
{k}}_{\parallel }\varepsilon _{t}(\omega ).
\end{equation}

For the dispersive model (\ref{bp}), the solution of
(\ref{SPPresonance}) yields a single branch of modes
$\omega_{\boldsymbol{\mathrm{k}}}$, which depends only on the angle
${\theta}$ of the wave vector with respect to the $x$-axis, not on
its magnitude,
\begin{equation}  \label{w_theta}
\omega_{\boldsymbol{\mathrm{k}}}=\omega_{\theta} = \frac{%
\omega_c}{2} ~ \mathrm{cos} ({\theta}) + \sqrt{ \frac{\omega_p^2%
}{2} + \frac{\omega_c^2}{4} (1 + \mathrm{sin}^2({\theta})) }.
\end{equation}

For $\omega_c>0$, one has $\omega_- < \omega_{\boldsymbol{\mathrm{k}}} < \omega_+ $%
, with
\begin{align}  \label{w+_w-}
& \omega_+ \equiv \omega_{{k}_x >0, {k}_y = 0} = \frac{1}{2}
\left( \omega_c + \sqrt{2\omega_p^2 + \omega_c^2} \right),  \nonumber \\
& \omega_- \equiv \omega_{{k}_x <0, {k}_y = 0} = \frac{1}{2} \left(
-\omega_c + \sqrt{2\omega_p^2 + \omega_c^2} \right) .
\end{align}

To have some insight into the physical meaning of the SPP resonance,
we numerically calculated the exact dispersion of the surface
plasmons using the formalism presented in Appendix
\ref{ApExactSPPs}. Figure \ref{figSPPs} depicts $\omega
_{{{\bf{k}}_{\parallel}}}^{{\rm{SPP}}}$ (the exact SPP dispersion)
versus $k_{\parallel}$ along different directions $\theta$ of the
wave vector. In each panel, the dashed horizontal line marks the SPP
resonance for which $\omega _{{{\bf{k}}_{\parallel}}}^{{\rm{SPP}}}
\to {\omega _\theta }$ with $\omega _\theta$ given by
(\ref{w_theta}). Thus, the quasi-static analytic solution determines
the SPPs with very short wavelengths (${{\bf{k}}_{\parallel}} \to
\infty$).

\begin{figure*}[tb!]
\begin{center}
\noindent \includegraphics[width=6.0in]{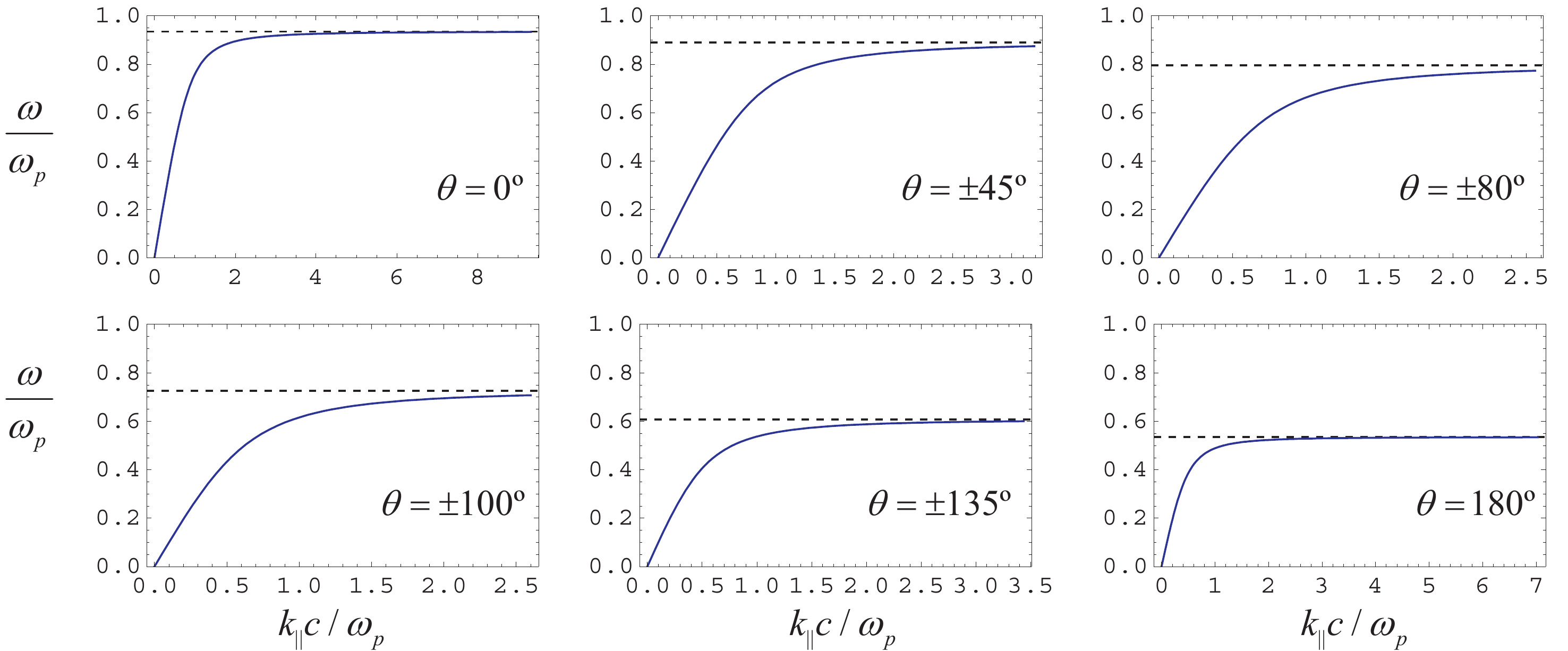}
\end{center}
\caption{Exact dispersion of the surface plasmons for different
angles $\theta$ of the wave vector. The dashed horizontal lines mark
the value of the quasi-static SPP resonance ($\omega_{\theta}$)
determined by (\ref{w_theta}). The cyclotron frequency is $\omega_c
= 0.4 \omega_p$. } \label{figSPPs}
\end{figure*}

In the limit of a vanishing bias field, $B_0 \to 0$, the
permittivity has a standard Drude-dispersion. In such a case, the
SPP resonance becomes angle independent,
\begin{equation}  \label{asympt}
{\lim _{{\omega _c} \to 0}}{\omega _{\mathbf{k}}} = \frac{{\omega
_p}}{\sqrt 2 },
\end{equation}
where ${\omega _p}/\sqrt 2$ is the frequency for which ${%
\varepsilon} = - 1$. The bias magnetic field shifts the SPP
resonance frequency and makes it direction-dependent. This creates
the opportunity to have light-matter interactions that depend
strongly on the direction of the emitted photons \cite{subPRL}.

\subsection{Optical force}

Next, we obtain an explicit expression for the optical force relying
on the modal expansion (\ref{FE1}) and on the quasi-static SPP
dispersion (\ref{w_theta}). It is shown that the light-matter
interactions are sculpted by the SPP resonances
$\omega=\omega_{\pm}$.

To begin with, we calculate the parameter ${A}_{\boldsymbol{{\rm
k}}_{\parallel }}$ in (\ref{elecpot2}) with the normalization
condition (\ref{norm}). This leads to
\begin{equation}
|{A}_{\boldsymbol{{\rm k}}_{\parallel }}|^{2}=\frac{2}{\varepsilon
_{0}}\left[ {{k}}_{\parallel }+\frac{\Lambda (\omega _{\theta
},\omega _{c},\omega _{p})}{2\tilde{{{k}}}_{\parallel }}\right]
^{-1}
\end{equation}%
where $\Lambda (\omega ,\omega _{c},\omega _{p})=\partial _{\omega
}\left(
\varepsilon _{t}\omega \right) \left( \tilde{{{k}}}_{\parallel }^{2}+%
{k}_{x}^{2}\right) +\partial _{\omega }\left( \varepsilon _{a}\omega
\right) {k}_{y}^{2}+\partial _{\omega }\left( \varepsilon _{g}\omega
\right) 2 {k}_{x}\tilde{{{k}}}_{\parallel }$.

Using $\boldsymbol{\mathrm{F}}_{n\mathbf{k}}\approx \left[ -\nabla
\phi _{\mathbf{k}}~~\boldsymbol{0}\right] ^{\mathrm{T}}$ and
(\ref{elecpot2}) in (\ref{FE1}), it is seen that the total force
(including both resonant and non-resonant components) may be
written as%
\begin{widetext}
\begin{align}
& \mathcal{F}_{j}=\rho _{ee}(t)\mathrm{Re}\left( \sum_{\omega _{\mathbf{k}%
}>0}\omega _{\mathbf{k}}\frac{|{A}_{\boldsymbol{\mathrm{k}}%
_{\parallel }}|^{2}}{S}e^{-2 {k}_{\parallel }d}\left( i\mathbf{k}%
_{\Vert }-k_{\Vert }\widehat{\mathbf{z}}\right) \boldsymbol{\gamma
}^{\ast }\cdot \left( i\mathbf{k}_{\Vert }-k_{\Vert
}\widehat{\mathbf{z}}\right) \left( -i\mathbf{k}_{\Vert }-k_{\Vert
}\widehat{\mathbf{z}}\right) \cdot \boldsymbol{\gamma
}\frac{1}{\omega _{\mathbf{k}}-\omega_0 -i0^{+}}\right)
\\
& +\left( 1-\rho _{ee}(t)\right) \mathrm{Re}\left( \sum_{\omega _{\mathbf{k}%
}>0}\omega _{\mathbf{k}}\frac{|{A}_{\boldsymbol{\mathrm{k}}%
_{\parallel }}|^{2}}{S}e^{-2 {k}_{\parallel }d}\left( i\mathbf{k}%
_{\Vert }-k_{\Vert }\widehat{\mathbf{z}}\right) \boldsymbol{\gamma
}\cdot
\left( i\mathbf{k}_{\Vert }-k_{\Vert }\widehat{\mathbf{z}}\right) \left( -i%
\mathbf{k}_{\Vert }-k_{\Vert }\widehat{\mathbf{z}}\right) \cdot \boldsymbol{%
\gamma }^{\ast }\frac{1}{\omega _{\mathbf{k}}+\omega_0
-i0^{+}}\right) . \nonumber
\end{align}%
To proceed, we use $\frac{1}{S}%
\sum\limits_{{\omega _{\mathbf{k}}}>0}\rightarrow \frac{1}{{{{\left( {2\pi }%
\right) }^{2}}}}\int {\int {d{k_{x}}d{k_{y}}}}$ to transform the
summation over the discrete modes into an integral. Moreover, using
polar coordinates ${{\bf{k}}_{\parallel}} = {k_{\parallel}}\left(
{\cos \theta ,\sin \theta ,0} \right)$, and noting that ${\omega
_{\bf{k}}} = {\omega _\theta}$, it is possible to write
\begin{align}
 &\mathcal{F}_{j}  = {\rho _{ee}}\left( t \right)\frac{{{{\left|
{\bf{\gamma }} \right|}^2}}}{{{\varepsilon _0}}}{\mathop{\rm
Re}\nolimits} \left\{ {\frac{1}{{{{\left( {2\pi }
\right)}^2}}}\int\limits_0^{2\pi } {d\theta } \int\limits_0^\infty
{d{k_{\parallel}}\,k_{\parallel}^2{\omega _\theta }{a_\theta }{e^{ -
2{k_{\parallel}}d}}{\Gamma _{ + ,\theta }}\,\frac{1}{{{\omega
_\theta } - \omega_0  - i{0^ + }}}\left( {i{{\bf{k}}_{\parallel}} -
{k_{\parallel}}{\bf{\hat
z}}} \right)} } \right\} \nonumber \\
& {\rm{ + }}\left( {1 - {\rho _{ee}}\left( t \right)}
\right)\frac{{{{\left| {\bf{\gamma }} \right|}^2}}}{{{\varepsilon
_0}}}{\mathop{\rm Re}\nolimits} \left\{ {\frac{1}{{{{\left( {2\pi }
\right)}^2}}}\int\limits_0^{2\pi } {d\theta } \int\limits_0^\infty
{d{k_{\parallel}}\,k_{\parallel}^2{\omega _\theta }{a_\theta }{e^{ -
2{k_{\parallel}}d}}{\Gamma _{ - ,\theta }}\,\frac{1}{{{\omega
_\theta } + \omega_0  - i{0^ + }}}\left( {i{{\bf{k}}_{\parallel}} -
{k_{\parallel}}{\bf{\hat z}}} \right)} } \right\}
\end{align}%
\end{widetext}
where we introduced
\begin{align} \label{Gamma_pm}
\Gamma _{+,\theta }& =\frac{1}{\left\vert \mathbf{\gamma
}\right\vert
^{2}k_{\Vert }^{2}}\left\vert \left( -i\mathbf{k}_{\Vert }-k_{\Vert }%
\widehat{\mathbf{z}}\right) \cdot \boldsymbol{\gamma }\right\vert ^{2}, \nonumber \\
\Gamma _{-,\theta }& =\frac{1}{\left\vert \mathbf{\gamma
}\right\vert
^{2}k_{\Vert }^{2}}\left\vert \left( -i\mathbf{k}_{\Vert }-k_{\Vert }%
\widehat{\mathbf{z}}\right) \cdot \boldsymbol{\gamma }^{\ast
}\right\vert ^{2},
\end{align}%
and ${a_\theta } \equiv {\left| {{A_{{{\bf{k}}_{\parallel}}}}}
\right|^2}{\varepsilon _0}k_{\parallel}^{}$, which are functions
only of $\theta$, not of ${k_{\parallel}}$. The integrals over
${k_{\parallel}}$ can be explicitly evaluated using
$\int\limits_0^\infty  {{e^{ - 2{k_{\parallel}}d}}k_{\parallel}^3}
dk_{\parallel} = \frac{3}{8}\frac{1}{{{d^4}}}$.

For the lateral force, only the two poles $\theta = \pm \theta_0$
for which the plasmon frequency matches the transition frequency of
the two-level atom (${\omega _{{\pm \theta_0}}=\omega_0}$)
contribute to the integral. In this case, we find that
\begin{align} \label{force_lateralQS}
\frac{\mathcal{F}_{x}}{\mathcal{F}_{0}} =-\rho _{ee}(t)  {\left.
{\frac{{{\omega _\theta }{a_\theta }\cos \theta }}{{\left|
{{\partial _\theta }{\omega _\theta }} \right|}}} \right|_{\theta  =
{\theta _0}}} \,
\frac{1}{2}%
\left( \Gamma _{+,\theta_0 }+\Gamma _{+,-\theta_0 }\right),
\nonumber \\
\frac{\mathcal{F}_{y}}{\mathcal{F}_{0}}=-\rho _{ee}(t) {\left.
{\frac{{{\omega _\theta }{a_\theta }\sin \theta }}{{\left|
{{\partial _\theta }{\omega _\theta }} \right|}}} \right|_{\theta  =
{\theta _0}}} \,
 \frac{1}{2}%
\left( \Gamma _{+,\theta_0 }-\Gamma _{+,-\theta_0 }\right).
\end{align}%
with
\begin{equation}
\mathcal{F}_{0}=\frac{3|\gamma |^{2}}{16\pi d^{4}\varepsilon _{0}},
\end{equation}%
a normalizing parameter with unities of force (N). Thus, in the
quasi-static approximation, the recoil force decays as $1/d^{4}$
with respect to the distance to the interface \cite{subPRL}.
Note that for a $\widehat{\mathbf{z}}$-directed dipole $\mathbf{\gamma }%
=\gamma \widehat{\mathbf{z}}$, we have $\Gamma _{+,\pm \theta }=1$.
As discussed in detail in \cite{subPRL}, since $a_\theta>0$ and
$\Gamma _{+,\pm \theta } \ge 0$ the sign of the force component
$\mathcal{F}_{x}$ (lateral force perpendicular to the bias magnetic
field) is independent of the dipole polarization and orientation.
Furthermore, the sign of $\mathcal{F}_{x}$ can be tuned with the
applied bias field. In contrast, the sign of $\mathcal{F}_{y}$
depends on the polarization state. It is highlighted that the
equation ${\omega _{{\pm \theta_0}}=\omega_0}$ has a solution only
if ${\omega _ - } \le {\omega _0} \le {\omega _ + }$. When
$\omega_0$ lies outside the frequency range of the SPP resonances
there are no poles, and the quasi-static approximation predicts a
vanishing lateral force. Indeed, plasmons with long wavelengths
interact weakly with the atom. Consistent with this, it is shown in
\cite{subPRL} that the exact lateral force quickly approaches zero
when ${\omega _0} < {\omega _-}$ or ${\omega _0} > {\omega _ +}$.

Equation (\ref{force_lateralQS}) reveals that the lateral force is
mainly determined by the plasmons that propagate with wave vector
directed along either $\theta=\theta_0$ or $\theta=-\theta_0$. As
further discussed in \cite{subPRL}, this implies that the momentum
transfer is determined by the canonical (Minkowski) momentum of
light, parallel to the wave vector, rather than by the kinetic
(Abraham) momentum, parallel to the Poynting vector (or
equivalently, to the group velocity).

The vertical component of the force is
\begin{align}
\frac{\mathcal{F}_{z}}{\mathcal{F}_{0}} & =-\rho _{ee}(t)\mathrm{Re}\left\{ \frac{1}{2\pi }%
\int_{0}^{2\pi }d\theta a_{\theta }\omega _{\theta }\Gamma _{+,\theta }\frac{%
1}{\omega _{\theta }-\omega_0-i0^+}\right\} \nonumber  \\
& -\left( 1-\rho _{ee}(t)\right) \mathrm{Re}\left\{ \frac{1}{2\pi }%
\int_{0}^{2\pi }d\theta a_{\theta }\omega _{\theta }\Gamma _{-,\theta }\frac{%
1}{\omega _{\theta }+\omega_0}\right\} . \label{FnormalQS}
\end{align}

In the steady-state limit, the only contribution to the normal force
is from the second integral, giving the Casimir-Polder force
 ($\mathcal{F}_C=\mathcal{F}_{z,t\rightarrow \infty }$) due to the vacuum
fluctuations at zero temperature,
\begin{equation}
\frac{\mathcal{F}_C}{\mathcal{F}_0}=- \frac{1}{2\pi }%
\int_{0}^{2\pi }d\theta a_{\theta }\omega _{\theta }\Gamma _{-,\theta }\frac{%
1}{\omega_0 +\omega _{\theta }} .
\end{equation}%
The Casimir-Polder force is clearly attractive ($\mathcal{F}_C<0$).
The sign of the dynamic normal force (Eq. (\ref{FnormalQS})) may be
either positive or negative.

\subsection{Limit of weak bias field}

\label{SectWeak}

It is interesting to further analyze the quasi-static solution in
the limit of a weak bias field $\omega_c \to 0$. Without loss of
generality, we suppose that the atom dipole moment is directed along
$z$, so that ${\Gamma _{ + , \pm {\theta _0}}}=1$. It can be shown
that for a weak bias magnetic field ${\omega _\theta } \approx
{\omega _{{\rm{spp}}}} + \frac{{{\omega _c}}}{2}\cos {\theta}$ and
$a_{\theta} \approx 1/2$, so that $|\partial_{\theta}
\omega_{\theta} |_{\theta = \theta_0} \approx  | \omega_c
\mathrm{sin} \theta_0|/2$ with $\omega_{\mathrm{spp}} =
\omega_p/\sqrt{2}$. Thus, the solution of ${\omega _{ \pm {\theta
_0}}} = {\omega _0}$ is such that $\cos {\theta _0} = 2\left(
{{\omega _0} - {\omega _{{\rm{spp}}}}} \right)/{\omega _c}$.
Therefore, for a weak bias and $\left| {{\omega _0} - {\omega
_{{\rm{spp}}}}} \right| < \left| {{\omega _c}} \right|/2$ the
nonzero-component of the lateral force (\ref{force_lateralQS})
reduces to
\begin{align}
\frac{\mathcal{F}_{x}}{\mathcal{F}_{0}}  &=-\rho_{ee}(t)
\frac{{{\omega _{{\rm{spp}}}}}}{{{\omega _c}}}\frac{{{\omega _0} -
{\omega _{{\rm{spp}}}}}}{{\sqrt {{{\left( {\omega _c^{}/2}
\right)}^2} - {{\left( {{\omega _0} - {\omega _{{\rm{spp}}}}}
\right)}^2}} }}. \label{ForceWeakLat}
\end{align}
Remarkably, as further discussed in \cite{subPRL}, the quasi-static
theory predicts that the lateral force diverges in the $\omega_c \to
0$ limit and for $\left| {{\omega _0} - {\omega _{{\rm{spp}}}}}
\right| = \left| {{\omega _c}} \right|/2$, i.e., when $\omega_0 =
\omega_+$ or $\omega_0 = \omega_-$. The lateral force vanishes when
$\left| {{\omega _0} - {\omega _{{\rm{spp}}}}} \right| > \left|
{{\omega _c}} \right|/2$.

Furthermore, for a weak bias and $\left| {{\omega _0} - {\omega
_{{\rm{spp}}}}} \right| < \left| {{\omega _c}} \right|/2$ the normal
force (\ref{FnormalQS}) is simply ${\mathcal{F}_{z}}= (1 -
\rho_{ee}(t)) {\mathcal{F}_{C}}$, with
\begin{align}
\frac{\mathcal{F}_{C}}{\mathcal{F}_{0}} & = - \frac{1}{2} \frac{
\omega_{\mathrm{spp}}  }{ \omega_{\mathrm{spp}} + \omega_0 }
\end{align}
the Casimir-Polder force when the atom is in the ground state. In
contrast, for $\left| {{\omega _0} - {\omega _{{\rm{spp}}}}} \right|
> \left| {{\omega _c}} \right|/2$ the normal force gains an
additional resonant component:
\begin{align}
\frac{\mathcal{F}_{z}}{\mathcal{F}_{0}}  =&-\rho_{ee}(t)
\frac{{{\omega _{{\rm{spp}}}}}}{2}\frac{{{\mathop{\rm sgn}} \left(
{{\omega _{{\rm{spp}}}}  - {\omega _0}} \right)}}{{\sqrt {{{\left(
{{\omega _0} - {\omega _{{\rm{spp}}}}}
\right)}^2} - {{\left( {{\omega _c}/2} \right)}^2}} }} \nonumber \\
& -(1 - \rho_{ee}(t)) \frac{1}{2}\,\frac{{{\omega
_{{\rm{spp}}}}}}{{{\omega _{{\rm{spp}}}} + {\omega _0}}}.
\label{ForceWeakNorm}
\end{align}
Hence, the quasi-static theory also predicts that the normal force
diverges when $\omega_0 = \omega_+$ or $\omega_0 = \omega_-$. In
contrast, the force component $\mathcal{F}_{C}$ has no resonances.
It will be shown in Sect. V that the force calculated with the
``exact'' Green function is finite when material loss and time
retardation are taken into account.

If the atom is prepared in an excited state, $\rho_{ee}(t=0)=1$, it
can be seen from (\ref{ForceWeakLat}) that the sign of the lateral
force ${\mathcal{F}_{x,t=0}}$ can be controlled either by changing
$\omega_c$ or $\omega_0$. In contrast, from (\ref{ForceWeakNorm}),
the sign of the normal force ${\mathcal{F}_{z,t=0}}$ only depends on
$\omega_0$, and hence cannot be dynamically tuned by flipping the
bias field.

\subsection{Influence of the atom polarization}

The atom polarization influences the strength of the Casimir-Polder
force through the non-negative coefficients $\Gamma _{\pm,\theta }$.
In particular, the lateral force ${\mathcal{F}_{x}}$ depends on the
sum $\Gamma _{+,\theta_0}+\Gamma _{+,-\theta_0}$.

From (\ref{Gamma_pm}) it is simple to check that ${\Gamma _{ +
,{\theta _0}}} = \frac{1}{{{{\left| {\boldsymbol{\gamma}}
\right|}^2}}}{\left| {\left( {i\cos {\theta _0}{\bf{\hat x}} + i\sin
{\theta _0}{\bf{\hat y}} + {\bf{\hat z}}} \right) \cdot
{\boldsymbol{\gamma}}} \right|^2}$, and thereby it is evident that
$0 \le {\Gamma _{ + ,{\theta _0}}} \le 2$. The maximum (${\Gamma _{
+ ,{\theta _0}}}=2$) is achieved for a polarization state such that
${\boldsymbol{\gamma}} \sim  { - i\cos {\theta _0}{\bf{\hat x}} -
i\sin {\theta _0}{\bf{\hat y}} + {\bf{\hat z}}}$.

The minimum ${\Gamma _{ + ,{\theta _0}}}=0$ is attained when
$\boldsymbol{\gamma}$ belongs to a two-dimensional complex vector
space generated by the complex vectors ${{\bf{v}}_{1,{\theta
_0}}}=i\cos {\theta _0}{\bf{\hat x}} + i\sin {\theta _0}{\bf{\hat
y}} + {\bf{\hat z}}$ and ${{\bf{v}}_{2,{\theta _0}}}=- \sin {\theta
_0}{\bf{\hat x}} + \cos {\theta _0}{\bf{\hat y}}$. Similarly, the
function $\Gamma _{+,-\theta_0}$ vanishes when that atom
polarization lies in the two-dimensional complex vector space
generated by the vectors ${{\bf{v}}_{1,{-\theta _0}}}$ and
${{\bf{v}}_{2,{-\theta _0}}}$. Thus, it follows that $\Gamma
_{+,\theta_0}+\Gamma _{+,-\theta_0}$ can be zero only when the atom
polarization vector is in the intersection of the two relevant
vector spaces, which can be shown to be the one-dimensional complex
vector space generated by $- {\bf{\hat x}} + i\cos {\theta
_0}{\bf{\hat z}}$. In other words, in the very special case in which
the atomic polarization state satisfies
\begin{equation}
{\boldsymbol{\gamma}} \sim  - {\bf{\hat x}} + i\cos {\theta
_0}{\bf{\hat z}}
\end{equation}%
the lateral force may vanish. This effect can be attributed to the
spin-momentum locking of the SPP \cite{Bliokh}. Note that by tuning
the bias magnetic field it is possible to adjust the value of
$\theta_0$ and thereby guarantee that the lateral force does not
vanish for any orientation of a given atom.

As an example, consider the case of a linearly polarized atom. Let
us introduce the polarization factor ${g_{{\boldsymbol{\gamma}}}} =
\frac{1}{2}\left( {{\Gamma _{ + ,{\theta _0}}} + {\Gamma _{ + , -
{\theta _0}}}} \right)$ which depends uniquely on the orientation of
the atom. If the atom has a random orientation the force is
determined by the orientational averaging of the polarization
factor, $\left\langle {{g_{{\boldsymbol{\gamma}}}}} \right\rangle$.
With the rough approximation $\left\langle
{{g_{{\boldsymbol{\gamma}}}}} \right\rangle \approx
\frac{1}{3}\left( {{g_{{\bf{\hat x}}}} + {g_{{\bf{\hat y}}}} +
{g_{{\bf{\hat z}}}}} \right)$, we find that $\left\langle
{{g_{{\boldsymbol{\gamma}}}}} \right\rangle  \approx \frac{2}{3}$. A
detailed analysis shows that this result is actually exact, i.e.,
the orientational averaging of the polarization factor for a
linearly polarized atom is precisely $\left\langle
{{g_{{\boldsymbol{\gamma}}}}} \right\rangle  = \frac{2}{3}$,
independent of the value of $\theta _0$.

\section{Numerical Examples}

\label{SectNumeric}

To illustrate the application of the developed theory, first we
discuss the validity of the quasi-static solution. In general, one
may expect that it should hold when the atom-interface distance is
much smaller than the wavelength ($d \ll 2 \pi c/\omega_p$ and $d
\ll 2 \pi c/\omega_0$), so that the effects of time retardation are
negligible. In addition, the quasi-static calculation assumes that
the material absorption is negligible. In all the numerical examples
presented below it is supposed that the dipole moment is along the
$z$-direction, so that $\mathcal{F}_y=0$.

\begin{figure*}[t]
\begin{center}
\noindent \includegraphics[width=5.5in]{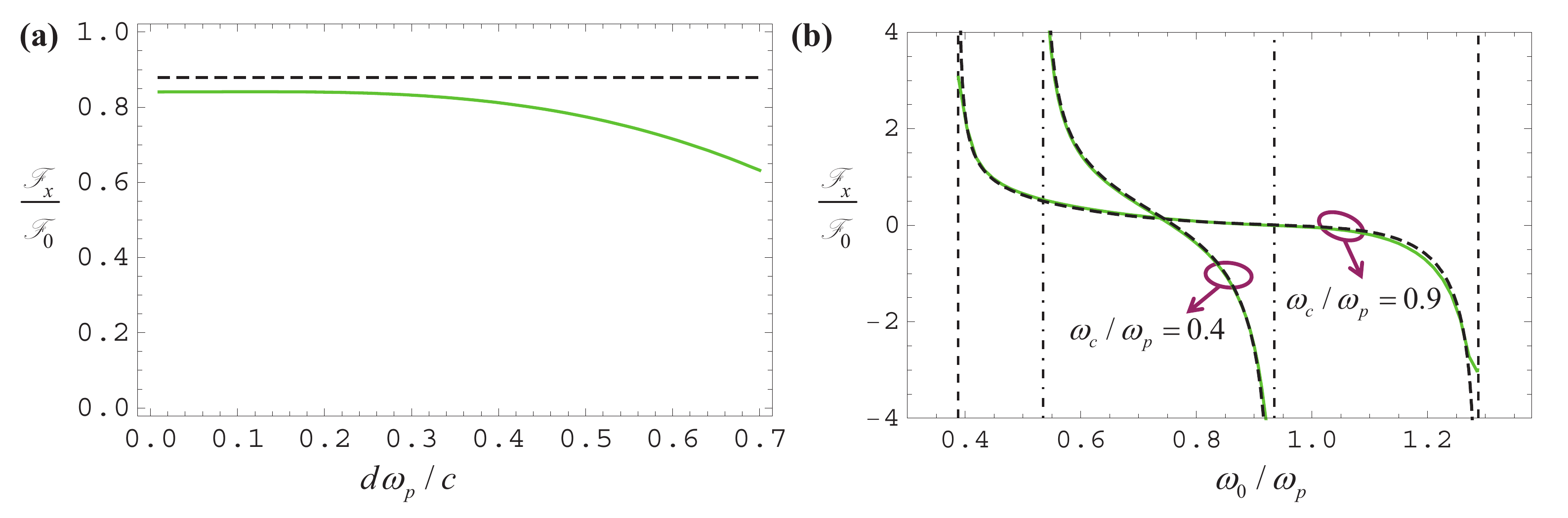}
\end{center}
\caption{ Comparison of the quasi-static (dashed black lines) and
exact (solid green lines) solutions for the lateral force for an
excited atom ($\rho_{ee}=1$). In the ``exact'' calculation the
plasma collision frequency is taken equal to $\Gamma=0.015
\omega_p$. (a) Lateral force as a function of the atom-interface
distance $d$ for a biased plasma with $\protect\omega_c /
\protect\omega_p = 0.4$ and an atom with $\protect\omega%
_0 / \protect\omega_p =0.65$. (b) Lateral force as a function of
$\protect\omega_0$ for the distance $d = 0.01 c/\protect\omega_p$.
The plot only shows the frequency range $\omega_- < \omega_0 <
\omega_+$, which for each case is delimited by the vertical
gridlines.} \label{figLF}
\end{figure*}

Figure \ref{figLF}(a) compares the exact solution for the lateral
force with the quasi-static approximation (\ref{force_lateralQS}),
showing how the normalized force varies with the distance to the
interface. For small distances $d \omega_p/c < 0.3$ the normalized
force is constant, confirming the $1/d^4$ power law. For larger
separations the quasi-static solution loses accuracy, and the force
follows a different power law. However, it should be noted that for
large $d$ the force is also much weaker (the coupling to the SPPs is
weaker) and hence it is not so relevant.

Figure \ref{figLF}(b) shows a comparison between the two calculation
methods when the distance is kept fixed and the atomic transition
frequency $\omega_0$ is varied. There is an excellent agreement
between the two solutions, further validating the quasi-static
approximation. The small discrepancy between the two methods for
$\omega \approx \omega_-$ and $\omega \approx \omega_+$ is
attributed in part to the fact that the exact calculation includes
the effect of material absorption ($\Gamma=0.015 \omega_p$). As
discussed in detail in \cite{subPRL}, the lateral force sign depends
on $\omega_0$. Furthermore, if the bias magnetic field is flipped
($\omega_c<0$) the sign of the lateral force is also flipped.

\begin{figure*}[t!]
\begin{center}
\noindent \includegraphics[width=5.5in]{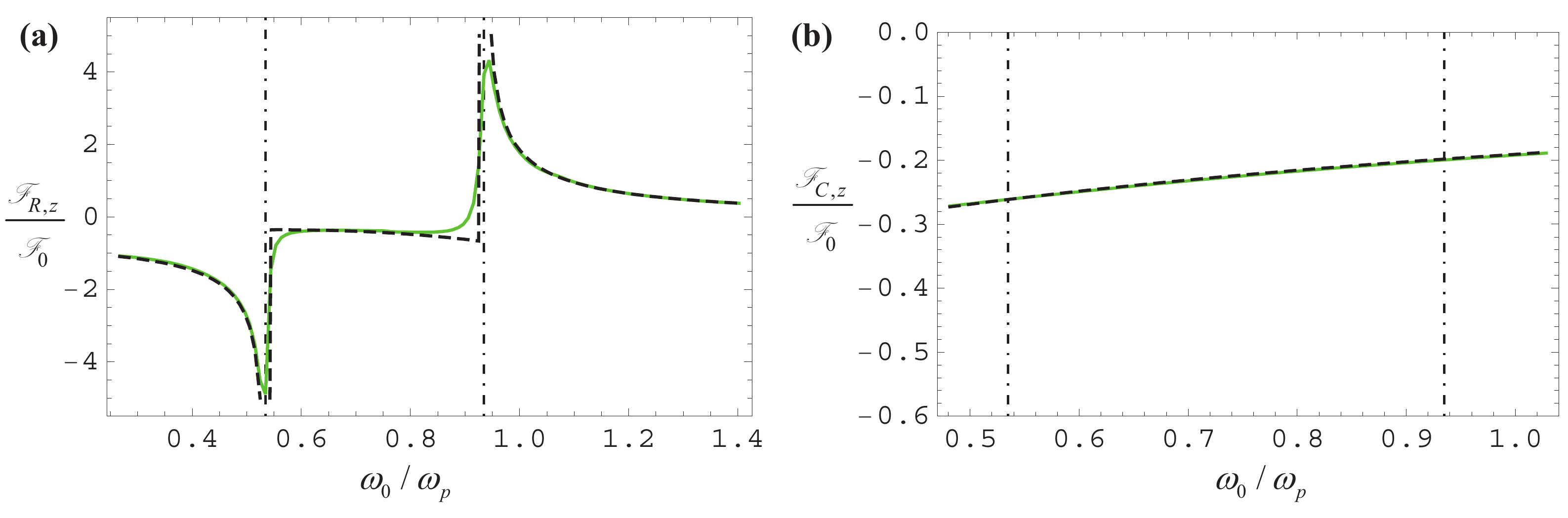}
\end{center}
\caption{Normal component of the (a) resonant force
${{\mathcal{F}}_{R}}$ and (b) the non-resonant force
${{\mathcal{F}}_{C}}$ as a function of the atom transition frequency
$\omega_0$ for $\omega_c = 0.4 \omega_p$ and $%
d = 0.01 c/\protect\omega_p$. Green solid lines: exact result for a
plasma collision frequency $\Gamma=0.015 \omega_p$. Black dashed
lines: quasi-static result. The vertical gridlines mark the points
$\omega = \omega_{\pm}$.} \label{figNF1}
\end{figure*}
\begin{figure*}[t!]
\begin{center}
\noindent \includegraphics[width=5.5in]{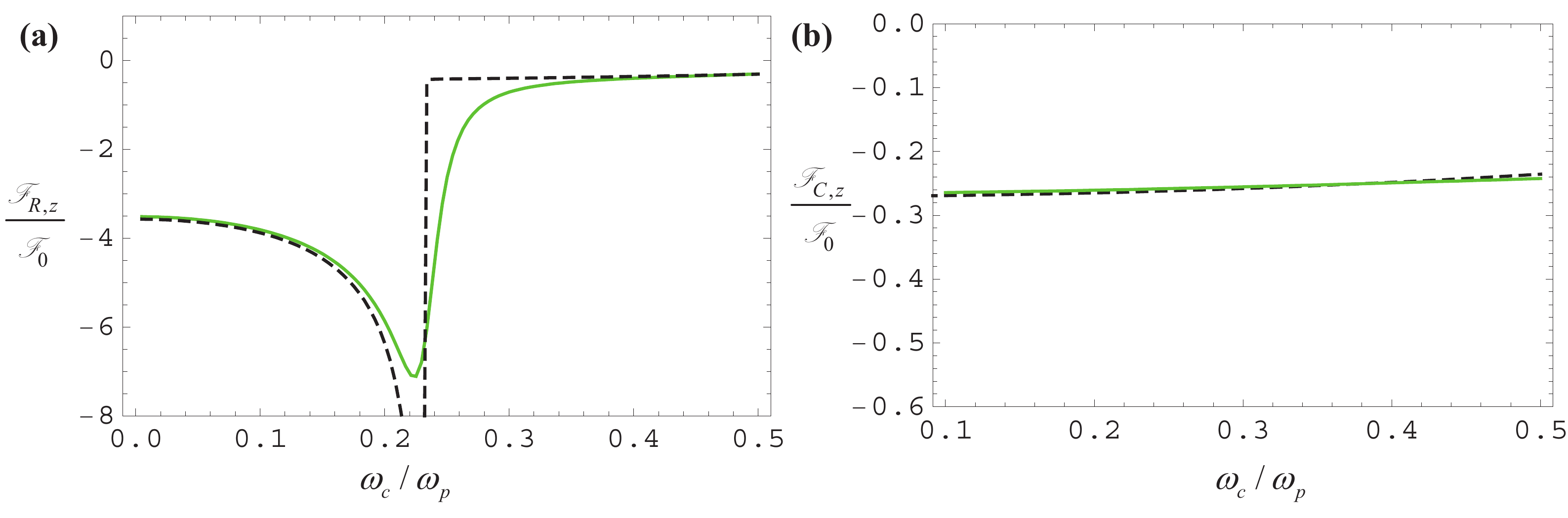}
\end{center}
\caption{Normal component of the (a) resonant force
${{\mathcal{F}}_{R}}$ and (b) the non-resonant force
${{\mathcal{F}}_{C}}$ as a function of the cyclotron frequency
$\omega_c$ for $\omega_0 = 0.6 \omega_p$ and $%
d = 0.01 c/\protect\omega_p$. Green solid lines: exact result for a
plasma collision frequency $\Gamma=0.015 \omega_p$. Black dashed
lines: quasi-static result.} \label{figNF2}
\end{figure*}

Next, we focus on the normal component of the force. Figure
\ref{figNF1} shows a comparison of the resonant part (panel-a) and
non-resonant part (panel-b) of the normal force calculated by the
exact and quasi-static solutions as a function of atom transition
frequency. Again, the quasi-static method agrees well with the exact
solution, excepting that the quasi-static solution diverges at
$\omega_{\pm}$ for the resonant part of the force, whereas the exact
calculation provides a finite result. Consistent with the discussion
in Sect. \ref{SectWeak}, the resonant force amplitude in the range
$\omega_- < \omega_0 < \omega_+$ is negligible as compared to the
value of the force outside this interval. Furthermore, the amplitude
of the non-resonant component of the force is typically at least one
order of magnitude smaller than the resonant component. As it also
happens for standard reciprocal materials, depending on the value of
$\omega_0$ the sign of resonant component ${{\mathcal{F}}_{R,z}}$
can be either positive or negative, but the nonresonant term is
always negative (attractive force). We would like to note that the
debate about the correct form of the metal response for low
temperatures and its implications on the thermal corrections of the
Casimir force \cite{thermal} does not affect our calculations of the
nonresonant force component, since the theory does not include any
thermal effects and gives the zero temperature limit solution.

Figure \ref{figNF2} illustrates how the normal force varies with the
plasma biasing strength. The non-resonant component of the force
${{\mathcal{F}}_{C}}$ is weakly sensitive to the magnetic bias.
Different from the lateral force (which has odd symmetry), the
vertical force components are even with respect to $\omega_c$ when
the atom is polarized along the vertical direction. Hence, it is not
possible to tune the sign of the normal force by changing the bias
magnetic field.

\begin{figure*}[tbh]
    \begin{center}
        \noindent \includegraphics[width=7in]{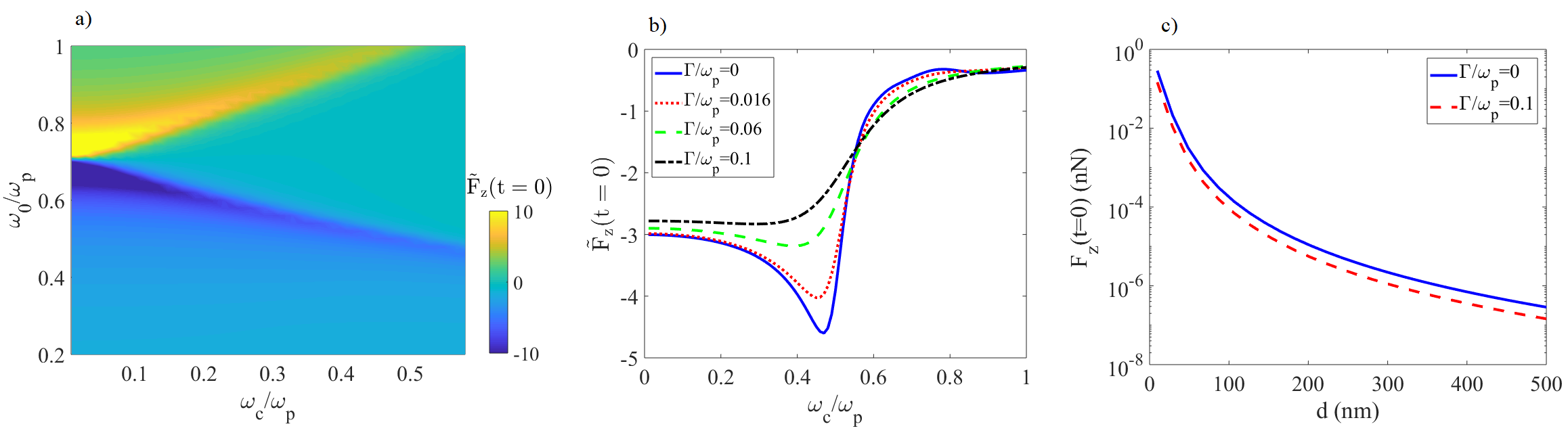}
    \end{center}
    \caption{(a) Density plot of the total normalized normal force
    ($\tilde{\mathrm{F}}_{z}=\mathcal{F}_{z}/\mathcal{F}_{0}$) as a function of atom transition frequency
    and bias strength for $ d = 0.01 c / \omega_p $.
    (b) Normal force as a function of the bias strength for an atom located at $ d=0.01c/\omega_p $ above
     a biased plasma (InSb-type material) with $ \omega_p / 2 \pi = 4.9$ THz and $ \omega_0/\omega_p = 0.5 $ for
     different values of collision frequency.
    (c) Normal force when the atom-interface distance is varied for $\omega_0/\omega_p = 0.93$ and $\omega_c / \omega_p = 0.4 $.} \label{Fz_InSb}
\end{figure*}
Figure \ref{Fz_InSb}(a) shows a density plot of the total normal
force (both resonant and non-resonant parts) for an excited atom
($\rho_{ee}=1$) as a function of the magnetic bias and of the atomic
transition frequency calculated by the exact solution. In this
panel, for a given cyclotron frequency $\omega_c$ the bright areas
correspond to $\omega_0=\omega_{\pm}$, where the peak of the normal
force occurs. It should be noted that for  $\omega_0 = \omega_+$ and
$\omega_0 = \omega_{-}$ the sign of the force changes. Furthermore,
consistent with (\ref{ForceWeakNorm}) it can be seen that a large
force is achievable at low bias.

Next, we use (\ref{bp}) as a simplified model of InSb with $\omega_p
/ 2\pi \approx 4.9$ THz, cyclotron frequency in the range of $0.25
\omega_p - \omega_p$ for a bias field of $1-4$ Tesla, and collision
frequency $ \Gamma/2\pi = 0.5$ THz \cite{Palik}. For simplicity, we
disregard the contribution of bound electrons to the permittivity
response of InSb, and in particular its static permittivity is taken
identical to unity. Figure \ref{Fz_InSb}(b) shows the effect of loss
on the total normal component of the force, when the collision
frequency varies from zero (lossless) to $0.5$ THz. Even in the
lossy case, there still exists considerable force applied to the
atom. In general, the effect of loss is relatively mild, but near
the resonant points $\omega_0 = \omega_{\pm}$ the system is more
sensitive to material absorption. Figure \ref{Fz_InSb}(c) shows the
normal force as a function of atom-InSb distance for lossless and
lossy InSb for a Rydberg atom \cite{RA} having $ \gamma = 7900 $ D
calculated by the exact solution. As seen, a significant force
persists even in the lossy case.

\section{Conclusion}

\label{SectConcl}

We have investigated the Casimir-Polder force on an excited atom in
a general (nonreciprocal, bianisotropic and dispersive)
electromagnetic environment under the Markov approximation. The
force is shown to have resonant and nonresonant contributions, and
we obtain explicit expressions for the quantum optical force in
terms of the system Green function. We have shown that a two-level
atom interacting with a topological gyrotropic material enables
exotic light-matter interactions, such as a lateral recoil force in
a laterally-invariant, homogeneous system with a sign independent of
the orientation and polarization of the atom \cite{subPRL}.
Furthermore, the strength of both the lateral and normal components
of the force is highly sensitive to the bias, and the sign of the
lateral force can also be externally controlled. In contrast, the
non-resonant Casimir-Polder force in the ground state is little
affected by the static magnetic field. To enable physical insight
into the phenomena, we have presented simple analytical expressions
for the quasi-static force, showing that surface plasmons play a
dominant role in these fluctuation induced forces. Remarkably, we
find that when the atomic transition frequency matches either the
SPP resonances $\omega_-$ or $\omega_+$ the resonant components of
the force (both lateral and normal) are greatly enhanced, and in the
quasi-static case (with no time retardation) and in the limit of no
material loss the force diverges. Furthermore, this phenomenon
persists even for a weak magnetic bias \cite{subPRL}. The material
absorption damps somewhat the strength of the force but the effect
appears to remain sufficiently strong to allow for an experimental
verification. We also present a Sommerfeld integral representation
for the Green function for a gyrotropic material half-space.

\appendix

\section{Modal Expansion of the Green function}

\label{ApGreen}

We introduce a frequency domain Green function $\boldsymbol{\mathrm{G}}(%
\boldsymbol{\mathrm{r}},\boldsymbol{\mathrm{r}}_{0})$ defined as a
six-tensor of the classical electric and magnetic dyadic Green
function,
\begin{equation} \label{Greendecomp}
\mathbf{G}=\left(
\begin{array}{cc}
\mathbf{G}_{\text{EE}} & \mathbf{G}_{\text{EH}} \\
\mathbf{G}_{\text{HE}} & \mathbf{G}_{\text{HH}}%
\end{array}%
\right)
\end{equation}
such that
\begin{equation} \label{Green}
{\boldsymbol{\mathrm{N}}}\cdot \boldsymbol{\mathrm{G}}=\omega \boldsymbol{%
\mathrm{M}}\cdot \boldsymbol{\mathrm{G}}+i\boldsymbol{\mathrm{I}}\delta (%
\boldsymbol{\mathrm{r}}-\boldsymbol{\mathrm{r}}_{0})
\end{equation}%
where $\boldsymbol{\mathrm{r}}$ is the observation point, $\boldsymbol{%
\mathrm{r}}_{0}$ is the source point, and
\begin{align}
& {\boldsymbol{\mathrm{N}}}=\left(
\begin{array}{cc}
\boldsymbol{0} & i\nabla \times \boldsymbol{\mathrm{I}}_{3\times 3} \\
-i\nabla \times \boldsymbol{\mathrm{I}}_{3\times 3} & \boldsymbol{0}%
\end{array}%
\right).
\end{align}%
The material matrix
$\mathbf{M}=\mathbf{M}\left({\bf{r}},\omega\right)$ determines the
electromagnetic properties of the environment, which in general may
be a bianisotropic nonreciprocal structure. In the limit of no loss,
it is possible to expand the
Green function into the natural eigenmodes $\boldsymbol{\mathrm{F}}_{n%
\boldsymbol{\mathrm{k}}}$ of the problem \cite{Silv1sm,
Mario_movingsm, SilvModalExpansions},
\begin{equation}
\boldsymbol{\mathrm{G}}(\boldsymbol{\mathrm{r}},\boldsymbol{\mathrm{r}}%
_{0},\omega )=\sum_{n\boldsymbol{\mathrm{k}}}\frac{i}{2(\omega _{n%
\boldsymbol{\mathrm{k}}}-\omega )}\boldsymbol{\mathrm{F}}_{n\boldsymbol{%
\mathrm{k}}}(\boldsymbol{\mathrm{r}})\otimes \boldsymbol{\mathrm{F}}_{n%
\boldsymbol{\mathrm{k}}}^{\ast }(\boldsymbol{\mathrm{r}}_{0}),
\end{equation}%
with $\boldsymbol{\mathrm{F}}_{n\boldsymbol{\mathrm{k}}}(\boldsymbol{\mathrm{%
r}}_{0})$ normalized as in Eq. (\ref{norm}). The sum is over all the
cavity
modes, i.e., modes with positive, negative and zero frequencies $\omega _{n%
\boldsymbol{\mathrm{k}}}$.

Taking into account that \cite{Mario_movingsm, SilvModalExpansions}
\begin{equation}
\sum_{n\boldsymbol{\mathrm{k}}}\frac{1}{2}\boldsymbol{\mathrm{F}}_{n%
\boldsymbol{\mathrm{k}}}(\boldsymbol{\mathrm{r}})\otimes \boldsymbol{\mathrm{%
F}}_{n\boldsymbol{\mathrm{k}}}^{\ast }(\boldsymbol{\mathrm{r}}^{\prime })=%
\boldsymbol{\mathrm{M}}_{\infty }^{-1}\delta
(\mathbf{r}-\mathbf{r}^{\prime }),
\end{equation}%
with ${\mathbf{M}_{\infty }}={\lim_{\omega \rightarrow \infty }}\mathbf{M}%
\left( {\mathbf{r},\omega }\right) $, it follows that
\begin{align}
& \boldsymbol{\mathrm{G}}=\sum_{n\boldsymbol{\mathrm{k}}}\frac{1}{2}\frac{%
i\omega _{n\boldsymbol{\mathrm{k}}}}{(\omega _{n\boldsymbol{\mathrm{k}}%
}-\omega )\omega }\boldsymbol{\mathrm{F}}_{n\boldsymbol{\mathrm{k}}}(%
\boldsymbol{\mathrm{r}})\otimes \boldsymbol{\mathrm{F}}_{n\boldsymbol{%
\mathrm{k}}}^{\ast }(\boldsymbol{\mathrm{r}}_{0})-\frac{i}{\omega }%
\boldsymbol{\mathrm{M}}_{\infty }^{-1}\delta (\boldsymbol{\mathrm{r}}-%
\boldsymbol{\mathrm{r}}_{0})  \nonumber \\
& ~~~=\sum_{\omega _{n\boldsymbol{\mathrm{k}}}>0}\frac{i\omega _{n%
\boldsymbol{\mathrm{k}}}}{2\omega }\left( \frac{1}{\omega _{n\boldsymbol{%
\mathrm{k}}}-\omega }\boldsymbol{\mathrm{F}}_{n\boldsymbol{\mathrm{k}}}(%
\boldsymbol{\mathrm{r}})\otimes \boldsymbol{\mathrm{F}}_{n\boldsymbol{%
\mathrm{k}}}^{\ast }(\boldsymbol{\mathrm{r}}_{0})\right.   \nonumber \\
& ~~~~~~+\left. \frac{1}{\omega _{n\boldsymbol{\mathrm{k}}}+\omega }%
\boldsymbol{\mathrm{F}}_{n\boldsymbol{\mathrm{k}}}^{\ast }(\boldsymbol{%
\mathrm{r}})\otimes \boldsymbol{\mathrm{F}}_{n\boldsymbol{\mathrm{k}}}(%
\boldsymbol{\mathrm{r}}_{0})\right) -\frac{i}{\omega }\boldsymbol{\mathrm{M}}%
_{\infty }^{-1}\delta
(\boldsymbol{\mathrm{r}}-\boldsymbol{\mathrm{r}}_{0}).
\end{align}%
In the second identity we used the fact that because of the reality
of the electromagnetic field the eigenmodes with negative
frequencies can be linked
to the eigenmodes with positive frequencies by a complex conjugation, $%
\boldsymbol{\mathrm{G}}^{\ast }(\boldsymbol{\mathrm{r}},\boldsymbol{\mathrm{r%
}}_{0},\omega )=\boldsymbol{\mathrm{G}}(\boldsymbol{\mathrm{r}},\boldsymbol{%
\mathrm{r}}_{0},-\omega ^{\ast })$, and we assume $\omega \in
\mathbb{R}$. For future reference, we decompose the Green function
as ${\bf{G}} = {{\bf{G}}^ + } + {{\bf{G}}^ - } + \frac{1}{{i\omega
}}{\bf{M}}_\infty ^{ - 1}\delta \left( {{\bf{r}} - {{\bf{r}}_0}}
\right)$, where
\begin{align}
{\left( {-i\omega }\right) \mathbf{G}}^{+}& =\sum\limits_{{%
\omega _{n\mathbf{k}}}>0}{\frac{{{\omega _{n\mathbf{k}}}}}{{2}}\frac{1}{%
\omega _{n\boldsymbol{\mathrm{k}}}-\omega }}\boldsymbol{%
\mathrm{F}}_{n\boldsymbol{\mathrm{k}}}(\boldsymbol{\mathrm{r}})\otimes
\boldsymbol{\mathrm{F}}_{n\boldsymbol{\mathrm{k}}}^{\ast }(\boldsymbol{%
\mathrm{r}}_{0})  \nonumber \\
{\left( {-i\omega}\right) \mathbf{G}}^{-}& =\sum\limits_{{%
\omega _{n\mathbf{k}}}>0}{\frac{{{\omega _{n\mathbf{k}}}}}{{2}}\frac{1}{%
\omega _{n\boldsymbol{\mathrm{k}}}+\omega }}\boldsymbol{%
\mathrm{F}}_{n\boldsymbol{\mathrm{k}}}^{\ast }(\boldsymbol{\mathrm{r}}%
)\otimes \boldsymbol{\mathrm{F}}_{n\boldsymbol{\mathrm{k}}}(\boldsymbol{%
\mathrm{r}}_{0}) \label{Gmp}
\end{align}%
are the positive and negative frequency parts of the Green
function, respectively.

\section{Green function for a gyrotropic material half-space}

\label{ApGreenHalfSpace}

Here, we derive the electric Green dyadic, ${{\bf{G}}_{{\rm{EE}}}}$,
for the case of a gyrotropic material half-space. As discussed in
Sect. \ref{SubSectLateral}, the vector
$\mathrm{\mathbf{E}}%
=-i\omega \boldsymbol{\mathrm{G}}_{\rm{EE}} \cdot
{\boldsymbol{\gamma }}$ gives the field emitted by a classical
dipole with electric dipole moment $\boldsymbol{\gamma }$. Hence,
${{\bf{G}}_{{\rm{EE}}}}$ can be found from the field radiated by a
generic dipole.

To calculate $\mathrm{\mathbf{E}}$, we note that the electromagnetic
field in the region $z>0$ (vacuum) is the
superposition of the primary field ($\mathbf{E^{\mathrm{p%
}}}$) and the scattered field ($\mathbf{E^{\mathrm{s%
}}}$). The primary field is given by ${{\bf{E}}^{\rm{p}}} = \left(
{\nabla \nabla  + k_0^2{\bf{\rm{I}}}} \right)\left(
\boldsymbol{\gamma }/\varepsilon _{0}\right) \Phi _{0}$ where
\begin{equation}
\Phi _{0}=\frac{e^{i {k}_{0}|\boldsymbol{\mathrm{r}}-\boldsymbol{%
\mathrm{r}}_{0}|}}{4\pi |\boldsymbol{\mathrm{r}}-\boldsymbol{\mathrm{r}}_{0}|%
}=\int \int d {k}_{x}d {k}_{y}\frac{e^{-\gamma _{0}|z-d|}}{%
2\gamma _{0}(2\pi )^{2}}e^{i({k}_{x}x+{k}_{y}y)}
\end{equation}%
is the Hertz potential, ${k}_{0}=\omega _{0}/c$, and $\gamma _{0}=%
\sqrt{{k_x}^{2}+{k_y}^{2}-{k}_{0}^{2}}$. Without loss of generality,
it is assumed that
the source point is $\boldsymbol{\mathrm{r}}%
_{0}=(0,0,d)$.

Following \cite{MarioPRXsm}, the scattered electric field above the
interface can be written as
\begin{equation}\label{Es_exact}
\mathbf{E}^{s}=\frac{1}{(2\pi )^{2}}\int \int d {k}_{x}d {k}%
_{y}e^{i\mathbf{k}_{\Vert }\cdot \mathbf{r}}\frac{e^{-\gamma _{0}(d+z)}}{%
2\gamma _{0}}\mathbf{C}\left( \omega ,\mathbf{k}_{\Vert }\right) \cdot \frac{%
\mathbf{\gamma }}{\varepsilon _{0}}
\end{equation}%
where
\begin{align} \label{CintAp}
& \mathbf{C}\left( \omega ,\mathbf{k}_{\Vert }\right)  \\
& =\left( \mathbf{I}_{\Vert }+\widehat{\mathbf{z}}\frac{i\mathbf{k}_{\Vert }%
}{\gamma _{0}}\right) \cdot \boldsymbol{\mathrm{R}}\left( \omega ,\mathbf{k}%
_{\Vert }\right) \cdot \left( i\gamma _{0}\mathbf{k}_{\Vert }\widehat{%
\mathbf{z}}+k_{0}^{2}\mathbf{I}_{\Vert }-\mathbf{k}_{\Vert }\mathbf{k}%
_{\Vert }\right)   \nonumber
\end{align}%
with $\mathbf{I}_{\Vert }=\widehat{\mathbf{x}}\widehat{\mathbf{x}}+\widehat{%
\mathbf{y}}\widehat{\mathbf{y}}$ and $\boldsymbol{\mathrm{k}}_{\parallel }=%
{k}_{x}\mathbf{\hat{x}}+{k}_{y}\mathbf{\hat{y}}$. Here, ${\mathbf{%
R}(\omega ,{k_{x}},{k_{y}})}$ is a $2\times 2$ reflection matrix
that relates the tangential (to the interface) components $x$ and
$y$ of the reflected electric field to the corresponding $x$ and $y$
components of the incident
field, $\left( {%
\begin{array}{ccccccccccccccccccc}
{E_x^{\rm s}}  \\
{E_y^{\rm s}}
\end{array}%
} \right) = \mathbf{R}\left( {\omega ,{k_x},{k_y}} \right) \cdot \left( {%
\begin{array}{ccccccccccccccccccc}
{E_x^{\rm inc}}   \\
{E_y^{\rm inc}}
\end{array}%
} \right)$ for the case of plane wave incidence. Since
(\ref{Es_exact}) holds for a generic electric dipole, it is
straightforward to verify that ${{\bf{G}}_{{\rm{EE}}}}$ has the
decomposition discussed in the main text, with the scattering part
of the Green function given by (\ref{GEE}).

To determine an explicit formula for $\mathbf{R}$, it is assumed
that the region $z < 0$ is filled with a gyrotropic material with
dielectric function given by (\ref{permittivity}). The incident
plane wave travels in the isotropic material region.

Evidently, the fields depend on $x$ and $y$ as $e^{i k_x x} e^{i k_y
y}$. In the region $z<0$ they can be written as a superposition of
two plane waves of the bulk gyrotropic medium with wave vector $\boldsymbol{%
\mathrm{k}}_i = \boldsymbol{\mathrm{k}}_{t,i} + {k}_y \mathbf{{\hat{y}%
}}$, with $\boldsymbol{\mathrm{k}}_{t,i} = {k}_x \mathbf{\hat{x}} +
{k}_{z,i} \mathbf{\hat{z}} $ ($i = 1,2 $). The subscript ``\emph{t}"
indicates that a certain vector component is perpendicular to the $y$%
-direction, which corresponds the direction of the bias magnetic
field. Setting ${k}_{z,i} = -i\gamma_{z,i}$ such that
$\mathrm{Re}\left( \gamma_{z,i} \right) > 0 $, the bulk mode
dispersion is \cite{HeatTransportsm}
\begin{widetext}
\begin{align}\label{gamma_z}
\gamma^2_{z,i} =& {k}^2_x - \frac{1}{2\varepsilon_t} \left[ \left(
\varepsilon_t \left(  \varepsilon_t + \varepsilon_a  \right)
-\varepsilon_g^2  \right) {k}^2_0 - \left(  \varepsilon_a +
\varepsilon_t  \right) {k}_y^2   \right] \notag \nonumber\\
& \pm \frac{1}{ 2 \varepsilon_t } \sqrt{  \left[  \left(
\varepsilon_t \left( \varepsilon_t + \varepsilon_a \right)
 - \varepsilon_g^2 \right) {k}_0^2 - \left( \varepsilon_a + \varepsilon_t \right) {k}_y^2   \right]^2  - 4\varepsilon_t \left[  \left( \varepsilon_t^2 - \varepsilon_g^2 \right) \varepsilon_a {k}_0^4 - 2 \varepsilon_t \varepsilon_a {k}_y^2 {k}_0^2 + \varepsilon_a {k}_y^4 \right]} .
\end{align}
\end{widetext}

Each of these possible solutions is associated with a plane wave.
For a plane wave superposition, the electric field is of the form \cite%
{HeatTransportsm}
\begin{align}  \label{E_gyro_bulk}
& \boldsymbol{\mathrm{E}} = \left( \Delta_1
\boldsymbol{\mathrm{k}}_1 \times \mathbf{\hat{y}} + \mathbf{k}_{t,1}
+ \theta_1 {k}_y \mathbf{\hat{y}}
\right) {A}_1 e^{ \gamma_{z,1}z }  \nonumber \\
& ~~~~~ + \left( \Delta_2 \boldsymbol{\mathrm{k}}_2 \times
\mathbf{\hat{y}}
+ \mathbf{k}_{t,2} + \theta_2 {k}_y \mathbf{\hat{y}} \right) {A%
}_2 e^{ \gamma_{z,2}z },
\end{align}
where the variation along $x$ and $y$ is omitted, ${A}_i~ (i=1,2) $
are expansion coefficients, and
\begin{align}
& \Delta_i = \frac{i\varepsilon_g {k}_0^2}{ {k}_0^2 \varepsilon_t -
( {k}_y^2 + {k}_{t,i}^2 ) }, ~~~ \theta_i =
\frac{ -{k}_{t,i}^2 }{ {k}_0^2 \varepsilon_a - {k}%
_{t,i}^2 }.
\end{align}
The magnetic field can be found from (\ref{E_gyro_bulk}), taking
into
account that for each plane wave $\boldsymbol{\mathrm{H}} = \boldsymbol{%
\mathrm{k}} \times \boldsymbol{\mathrm{E}} / \omega \mu_0 $. Using (\ref%
{E_gyro_bulk}) and the magnetic field expression, the expansion
coefficients can be eliminated, leading to
\begin{align}
\left(%
\begin{array}{c}
\eta_0 {H}_y \\
-\eta_0 {H}_x%
\end{array}%
\right) = - \boldsymbol{\mathrm{Y}}_g \cdot \left(%
\begin{array}{c}
{E}_y \\
{E}_x%
\end{array}%
\right),
\end{align}
with $\eta_0$ the vacuum impedance and
\begin{align}
&\boldsymbol{\mathrm{Y}}_g = \left(%
\begin{array}{cc}
\frac{\Delta_1 {k}_{t,1}^2 } {{k}_0} & \frac{\Delta_2 {k%
}_{t,2}^2 } {{k}_0} \\
\frac{\Delta_1 {k}_x {k}_y + i \gamma_{z,1} ( \theta_1 -1 ) {k}_y}{
{k}_0 } & \frac{\Delta_2 {k}_x {k}_y + i
\gamma_{z,2} ( \theta_2 -1 ) {k}_y}{ {k}_0 }%
\end{array}%
\right)   \nonumber \\
& ~~~~~~~~ \cdot \left(%
\begin{array}{cc}
{k}_x + i \gamma_{z,1} \Delta_1 & {k}_x + i \gamma_{z,2}
\Delta_2 \\
\theta_1 {k}_y & \theta_2 {k}_y%
\end{array}%
\right)^{-1}.
\end{align}
Note that the considered field distribution corresponds to a wave
that propagates towards the $-z$ direction in the gyrotropic
material. Likewise, it is possible to show that for a wave that
propagates in the isotropic
dielectric (air region) in the $\pm z$ direction the fields satisfy $\left( {%
\begin{array}{ccccccccccccccccccc}
{{\eta _0}{H_y}}  \\
{\ - {\eta _0}{H_x}}
\end{array}%
} \right) = \pm {\mathbf{Y}_0} \cdot \left( {%
\begin{array}{ccccccccccccccccccc}
{E_x}   \\
{E_y}
\end{array}%
} \right)$ with
\begin{align}
& \boldsymbol{\mathrm{Y}}_0 = \frac{1}{ i {k}_0 \gamma_0 } \left(
\begin{array}{cc}
-\gamma_0^2 + {k}_x^2 & {k}_x {k}_y \\
{k}_x {k}_y & -\gamma_0^2 + {k}_y^2%
\end{array}%
\right),
\end{align}
where $\gamma _0^2 = k_x^2 + k_y^2 - k_0^2$.

It is now straightforward to obtain the reflection matrix $\boldsymbol{%
\mathrm{R}}$. Noting that the field in the region $z>0$ is a
superposition of the incident and reflected waves and that the field
in the region $z<0$ is of the form (\ref{E_gyro_bulk}), it follows,
imposing the continuity of the tangential fields at the interface,
that $\boldsymbol{\mathrm{Y}}_0
\cdot ( - \mathbf{1} + \boldsymbol{\mathrm{R}}) = - \boldsymbol{\mathrm{Y}}%
_g ( \mathbf{1} + \boldsymbol{\mathrm{R}})$. From here we obtain the
desired result,
\begin{equation}  \label{R_matrix}
\boldsymbol{\mathrm{R}} = \left(\boldsymbol{\mathrm{Y}}_0 + \boldsymbol{%
\mathrm{Y}}_g\right)^{-1} \cdot \left( \boldsymbol{\mathrm{Y}}_0 -
\boldsymbol{\mathrm{Y}}_g \right).
\end{equation}

\section{Exact dispersion equation for the surface plasmons}

\label{ApExactSPPs}

Here, we derive the exact dispersion equation for the surface
plasmons assuming that the region $z>0$ is free-space and that the
region $z<0$ is a magnetized gyrotropic plasma. It is supposed that
the interface between the two regions is perfectly smooth, and hence
possible contributions to the SPPs dispersion due to surface
roughness are disregarded.

The fields in the two regions can be expanded into evanescent plane
waves. In particular, in the bulk gyrotropic medium the modes can be
written as a superposition of two plane waves with $z$ propagation
factor defined as in (\ref{gamma_z}) and the electric field given by
(\ref{E_gyro_bulk}). The associated magnetic field in the
region $z<0$ can be found using $\boldsymbol{\mathrm{H}} = \boldsymbol{%
\mathrm{k}} \times \boldsymbol{\mathrm{E}} / \omega \mu_0$ for each
plane wave term.

The fields in the vacuum region ($z>0$) can be expanded as
\begin{align}
& \boldsymbol{\mathrm{E}} = - \left[ {B}_1 \boldsymbol{\mathrm{k}}_0
\times \mathbf{\hat{z}} + {B}_2 \boldsymbol{\mathrm{k}}_0 \times (
\boldsymbol{\mathrm{k}}_0 \times \mathbf{\hat{z}} ) \right]
e^{-\gamma_0 z}
\nonumber \\
& \omega \mu_0 \boldsymbol{\mathrm{H}} = - \left[ {B}_1 \boldsymbol{%
\mathrm{k}}_0 \times ( \boldsymbol{\mathrm{k}}_0 \times
\mathbf{\hat{z}} ) -
{B}_2 \frac{\omega^2}{c^2} ( \boldsymbol{\mathrm{k}}_0 \times \mathbf{%
\hat{z}} ) \right] e^{-\gamma_0 z}
\end{align}
with $\boldsymbol{\mathrm{k}}_0 = {k}_x \mathbf{\hat{x}} + {k}%
_y \mathbf{\hat{y}} + i \gamma_0 \mathbf{\hat{z}} $ and $\gamma_0 =
\sqrt{ {k}_x^2 + {k}_y^2 - \omega^2 / c^2 } $. By matching the
tangential electromagnetic fields at the interface ($z=0$) we arrive
at the
following system of equations, 
\begin{align}
&\left(%
\begin{array}{cccc}
{k}_x + i \gamma_{z,1}\Delta_1 & {k}_x + i \gamma_{z,2} \Delta
& {k}_y & {k}_x i \gamma_0 c/\omega \\
\theta_1 {k}_y & \theta_2 {k}_y & -{k}_x & {k}%
_yi\gamma_0 c/\omega \\
\Phi_1 & \Phi_2 & {k}_x i \gamma_0 & - {k}_y \omega / c \\
-\Delta_1 {k}^2_{t,1} & -\Delta_2 {k}^2_{t,2} & {k}_y i
\gamma_0 & {k}_x \omega / c%
\end{array}%
\right) \cdot  \nonumber \\
& \left(%
\begin{array}{c}
{A}_1 \\
{A}_2 \\
{B}_1 \\
{B}_2 \frac{\omega }{c}%
\end{array}%
\right) = \boldsymbol{0}_{4 \times 1}
\end{align}
where $\theta_i$, $\Delta_i$, and $\gamma_{z,i} $ are defined in
Appendix \ref{ApGreenHalfSpace} and $\Phi_i = \Delta_i {k}_x {k}_y +
i \gamma_{z,i} (\theta_i -1 ) {k}_y $, $(i=1,2)$. Setting the
determinant of the matrix equal to zero leads to the exact SPP
dispersion equation.

\begin{acknowledgments}
 The authors gratefully acknowledge
discussions with S. Buhmann. This work was partially funded by
Funda\c{c}\~{a}o para a Ci\^{e}ncia e a Tecnologia under project
PTDC/EEI-TEL/4543/2014 and by Instituto de Telecomunica\c{c}\~{o}es
under project UID/EEA/50008/2013. M. S. thanks the CNRS and the
group Theory of Light-Matter and Quantum Phenomena of the
Laboratoire Charles Coulomb for hospitality during his stay in
Montpellier.
\end{acknowledgments}


\begin{thebibliography}{99}
\bibitem{footnoteLoss} The decay rate $\Gamma_{eg}$
is not included in the formula but its effect is negligible because
the integration is over the imaginary axis and for typical atomic
systems $\Gamma_{eg}\ll \omega_0$.
\end{thebibliography}
\end{document}